
\input amstex
\documentstyle{amsppt}
\def\C{\Cal C}
\def\A{\Cal A}
\def\B{\Cal B}
\def\F{\Cal\Phi}
\def\Q{\Cal Q}
\def\a{\alpha}
\def\e{\varepsilon}
\def\th{\theta}
\def\l{\lambda}
\def\s{\sigma}
\def\om{\omega}
\def\Om{\Omega}
\def\g{\gamma}
\def\be{\beta}
\def\d{\delta}
\def\D{\Delta}
\def\ot{\otimes}
\def\lra{\longrightarrow}
\def\w{\wedge}
\def\G{\Cal G\text{\it r}(\Cal C)}
\def\de{d_\phi}
\def\M{\Omega^*(A,\phi )}
\def\dek{d_{\phi^k}}
\def\O#1#2{\Omega^{#1}(#2)}
\def\pa{\partial}
\def\N#1{\Cal N_{#1}(A,\phi )}
\def\L#1{\Cal L_{#1}}
\def\ls#1{\Lambda_{\hat\sigma}^{#1}(M)}
\def\t{\bullet}
\def\Ni#1{\Cal N_{#1}(M)}
\def\n{\nabla}
\def\dq{\underset{q}\to\cdot}
\def\cq{\underset{q}\to\circ}
\hyphenation{com-mu-ta-ti-ve}
\hyphenation{fol-low-ing}
\hyphenation{con-si-de-ra-ti-ons}
\hyphenation{as-su-me}
\hyphenation{mo-ti-va-te}
\TagsOnRight
\topmatter
\title Calculus and Quantizations\\ over Hopf algebras\endtitle
\author V.Lychagin\endauthor
\affil
International Sophus Lie Center, Moscow, Russia \& Center for Advanced
Study at the Norwegian Academy of Science and Letters\endaffil
\address
P.Box 546,119618, Moscow, Russia
\endaddress
\curraddr
P.O.Box 7606, Skillebekk 0205, Oslo, Norway
\endcurraddr
\email
Valentin.Lychagin\@ shs.no
\endemail
\abstract
In this paper we outline an approach to calculus over quasitriangular Hopf
algebras. We construct braided differential operators  and introduce a general
notion of quantizations in monoidal categories . We discuss some
applications to quantizations of differential operators.
\endabstract
\endtopmatter
\document
\head {\bf 0.Introduction}\endhead
In this paper we study differential operators in the framework of monoidal
categories equipped with a braiding or symmetry. To be more concrete, we
choose as an example the category of modules over quasitriangular Hopf
algebra.

We introduce (braided) differential operators in a purely algebraic manner.
This gives us a possibility to develop calculus in an intrinsic way
without enforcing any type of Leibniz rule.

A general notion of quantization in monoidal categories, proposed in this
paper, is a natural isomorphism of the tensor product bifunctor equipped with
some natural coherence conditions. The quantization "deforms" all natural
algebraic and differential objects in the monoidal category.

There are now a number of different approaches to the construction of a
 calculus: the universal construction for associative algebras [C],[K],[DV],
fermionic and colour calculus [JK],[BMO],[KK], the calculus for quadratic
algebras [WZ],[M], covariant calculus on Hopf algebras [W], etc. Here we
would like to illustrate the general scheme for differential calculus
suggested in [L1],[L2] on the example of the monoidal category of modules over
quasitriangular Hopf algebra.

The paper is organized as follows. In section 1 we build up modules of
(braided) differential operators in the category of modules over
quasitriagular Hopf algebra. In section 2 we consider (braided) derivations
as a special type of 1-st order (braided) differential operators. We show that
these operators may be described by a braided Leibniz rule. We
should note that for general braidings (not symmetries) the rule consists of
four identities.

As usual we introduce braided differential 1-forms $\Om^1(A)$ as a
representative object for the functor of braided derivations in new category
with morphisms generated by braided differential operators of degree 0.

Our construction of braided differential forms and the de Rham complex is
based on the following two assumptions:
\roster
\item An algebra of braided differential forms should be a braided commutative
algebra generated by the base algebra $A$ and symmetric bimodule $\Om^1(A)$,and
\item de Rham differential $d$ should be a braided derivation of the algebra,
such that $d^2=0.$
\endroster
To analyze the first condition we describe all braidings in the category of
$\Bbb Z$--graded objects over a given monoidal category and show that the
second condition defines a special class of braidings which we call
differential
 prolongations of the given one.

It is an experimental fact that modules of braided derivations (in various
definitions) have no good Lie algebra structure. We remark that modules of
braided differential forms may be considered as Lie coalgebras and condition
$d^2=0$ may be considered as the analogue of co- Jacobi identity.

In section 3 we introduce a quantization of functors acting between monoidal
categories. In our definition, a quantization is a natural isomorphism equipped
with some natural coherence conditions and the functor considered together with
a quantization is simply a monoidal functor [cf.McL, Ep].

We suggest two ways for calculation of quantizations. One of them reduces the
calculation to nonlinear cohomologies. The other describes
quantizations in terms of multiplicative Hochschild cohomologies of the
Grothendieck ring of the given monoidal category. These constructions are
illustrated by some examples. Thus for the monoidal category of
representations of torus the quantizers may be described in terms of invariant
Poisson structures. Their construction given in 3.6. produces the Moyal
quantizations[BFFLS,V]. In the same way we obtain series of quantizers for
categories of representations of compact Lie groups.

This paper was written during a visit to the Centre for Advanced Study at the
Norwegian Academy of Science and Letters. I would like to thank Prof.A.Laudal
for hospitality. It is also my pleasure to thank Profs. D.Gurevich, A.Laudal,
A.Sletsj\o e  for valuable discussions.
\head
{\bf 1.Braided differential operators over quasitriangular Hopf algebras}
\endhead
{\bf 1.1.} Let $k$ be a commutative ring with unit. We shall assume that all
$k$--algebras under consideration have a unit, and that all algebra
homomorphisms are unit--preserving.

Let $H$ be a Hopf $k$--algebra with a coproduct $\D :H\lra H\ot H,$ counit
$\e :H\lra k$ and antipode $S:H\lra H.$

Denote by $\C=\Cal Mod_H$ the category of left $H$--modules. Morphisms in
this category are $H$--module homomorphisms.

To convert $\C$ into a monoidal category, we define a bifunctor of tensor
product
$$\ot :\C\times\C\lra\C$$
to be the usual tensor product of modules over $k:$
$X\ot Y=X\underset k\to\ot Y.$

We define an $H$--module structure in the tensor product as follows:
$$h(x\ot y)=\sum_h h_{(1)}(x)\ot h_{(2)}(y),$$
where $x\in X,y\in Y,h\in H$ and $\D (h)=\sum_h h_{(1)}\ot h_{(2)}$ in the
Sweedler notations [S].

The bifunctor of internal homomorphisms $X,Y\mapsto Hom(X,Y)$ in the monoidal
category is the adjoint bifunctor for the tensor product bifunctor.

In our case $Hom(X,Y)$ coincides with the module  of $k$--morphisms

$Hom_k(X,Y)$ equipped with the following $H$--module structure :
$$h(f)(x)=\sum_h h_{(1)}\cdot f(S(h_{(2)})( x)),$$
where $h\in H,x\in X,f\in Hom(X,Y).$

{\bf 1.2.}By the usual definition of algebras in monoidal categories algebras
in the category $\C$ are $H$--module algebras, i.e. $k$--algebras $A$
which are: (1) $H$--modules, (2) multiplication maps $\mu :A\ot A\lra A,\quad
\mu (a\ot b)=a\cdot b,$ are morphisms in $\C$.

The last condition means that
$$h(a\cdot b)=\sum_h h_{(1)}(a)\cdot h_{(2)}(b),$$
and
$$h(1)=\e (h),$$
for all $a,b\in A,\quad h\in H.$

A left $A$--module $P$ in the category $\C$ is an $H$-- and $A$-- module
$$\mu^l:A\ot P\lra P,\quad\mu^l (a\ot p)=a\cdot p,$$
such that
$$h(a\cdot p)=\sum_h (h_{(1)}(a))\cdot (h_{(2)}(p))$$
for all $a\in A,\quad p\in P.$

In the same way one defines right $A$--modules and $A-A$ bimodules in the
category $\C$.

{\bf 1.3.} Recall [JS] that a {\it braiding} in a monoidal category $\C$ is a
natural isomorphism
$$\s_{X,Y} :X\ot Y\lra Y\ot X,\qquad\forall X,Y\in\Cal Ob(\C ),$$
such that the following hexagon conditions hold:
$$\left\{\aligned
&\s_{X\ot Y,Z}=\bigl(\s_{X,Z}\ot id_Y\bigr)\circ\bigl( id_X\ot\s_{Y,Z}\bigr)
,\\
&\s_{X,Y\ot Z}=\bigl( id_Y\ot\s_{X,Z}\bigr)\circ\bigl(\s_{X,Y}\ot id_Z\bigr)
.\\
\endaligned\right.$$

It is easy to show that any braiding in the category is given by the
{\it braiding element} $\s\in H\ot H,$ such that the above hexagon conditions
$$\left\{\aligned
&\bigl( id_H\ot\D\bigr) (\s )=(\s\ot 1)\cdot\s_{13},\\
&\bigl(\D\ot id_H\bigr) (\s)=(1\ot\s )\cdot\s_{13},\\
\endaligned\right.\tag1$$
and the condition that $\s$ is a $\C$--morphism
$$\s\cdot\tau ((\D (h))=\D (h)\cdot\s\tag2$$
hold for all $h\in H.$

Here we denote by
$$\tau :H\ot H\lra H\ot H$$
the natural {\it twist} $\quad\tau (a\ot b)=b\ot a.$

As usual, $\s_{1,3}$ denoted the element of $H\ot H\ot H$ which is $\s$ in
the 1-st and 3-rd factors, i.e. $\s_{13}=\sum \s '\ot 1\ot\s '',$ if
$\s =\sum\s '\ot\s '',$ or $\s_{13}=(id_H\ot\tau )(\s\ot 1).$

The definition of the braiding by means of the braiding element is the
following:
$$\s_{X,Y}(x\ot y)=\s\cdot (y\ot x)=\sum\s '(y)\ot\s ''(x),$$
for all $x\in X,y\in Y.$

Following Drinfeld [D] a Hopf algebra $H$ equipped with a braiding element
$\s$ is called a {\it quasitriangular Hopf algebra.}

In a quasitriangular Hopf algebra one has the following relations on the
braiding element $\s$:
$$\s_{12}\s_{13}\s_{23}=\s_{23}\s_{13}\s_{12},\quad\text{(Quantum Yang--Baxter
equation)}$$
where $\s_{12}=\s\ot 1,\s_{23}=1\ot\s ,\quad$ and
$$\left\{\aligned
&(S\ot id)(\s )=(id\ot S)(\s )=\s^{-1},\quad (S\ot S)(\s )=\s ,\\
&(\e\ot id)(\s )=(id\ot\e )(\s )=1.
\endaligned\right.$$
A braiding $\s$ is called a {\it symmetry} [McL] if
$\s_{Y,X}\circ\s_{X,Y}=id_{X,Y},$
or
$\s\cdot\tau (\s )=1$
in terms of braiding elements.

A quasitriangular Hopf algebra $(H,\s )$ is called a {\it triangular Hopf
algebra} if $\s$ is a symmetry.
\example{Examples}

(1) Let $G$ be a finite multiplicative group and $H=k(G)$ be a $k$--algebra of
functions on $G$ with values in $k$.

Define a coproduct, a counit an and antipode as usual:
$$(\D f)(x,y)=f(xy),\quad\e (f)=f(e),\quad S(f)(x)=f(x^{-1}),$$
where $f\in k(G),\quad x,y\in G,\quad e$ is a unit of $G$ and we identify
$k(G)\ot k(G)$ with $k(G\times G).$

Let $\th_x,\quad x\in G$ be a basis in $k(G)$ such that:$\th_x(x)=1,$ and
$\th_x(y)=0,$ if $x\ne y.$

In terms of this basis we have:
\roster
\item $\th_x\cdot\th_x=\th_x,$ and $\th_x\cdot\th_y=0,$ if $x\ne y,$
\item $\D\th_x=\sum_{y\in G}\th_y\ot\th_{y^{-1}x},$
\item $\e (\th_x)=0,$ if $x\ne e,$ and $\e (\th_e)=1,$
\item $S(\th_x)=\th_{x^{-1}}.$
\endroster
Property (1) shows that $\th_x$ are projectors and therefore the category
$Mod_H$ coincides with a category of $G$--graded modules, $X=\sum_{g\in G}X_g,$
equipped with the following $H$--action:
$$\th_g(\sum_{h\in G}x_h)=x_g.$$
Morphisms $f:X\lra Y$ in the category are $G$--graded $k$--morphisms:
$f(X_g)\subset Y_g.$

Tensor product $X\ot Y$ in this category is the usual $G$--graded product:
$$(X\ot Y)_g=\sum_{h\in G}X_h\ot Y_{h^{-1}g}.$$
Internal homomorphisms $Hom(X,Y)$ coinside with modules $k$--homomorphisms
equipped with the grading:
$$Hom(X,Y)_g=\{ f\in Hom_k(X,Y)\vert f(X_h)\subset Y_{gh},\forall h\in G\} .$$
A braiding element $\s\in H\ot H$ can be written down as follows:
$$\s =\sum_{a,b\in G}\s (a,b)\,\th_a\ot\th_b.\tag1$$
{}From the hexagon axioms we obtain the following conditions on the function
$\s (a,b):$
$$ \s(ab,c)=\s (a,c)\cdot\s (b,c),\tag2$$
and
$$\s (a,bc)=\s (a,b)\cdot\s (a,c),\tag3$$
for all $a,b,c\in G.$

But condition 1.3.(2) holds if and only if $G$ is an Abelian group.

Therefore, any braiding in the category of $G$--graded modules over an Abelian
group $G$ is given by the group bihomomorphism $$\s :G\times G\lra U(k),$$
where $U(k)$ is the unit group of the ring $k$.

The braiding is a symmetry if and only if the following multiplicative skew
symmetry property holds: $\s (a,b)\s (b,a)=1.$

(2) Let $G$ be a finite group and $H=k[G]$ the group algebra, $k[G]=(k(G))^*.$

Denote by $\d_g,\quad g\in G$ the dual basis of $\d$--functions:
$\d_g(\th_g)=1,$ and $\d_g(\th_h)=0,$ if $g\ne h.$

In terms of this basis a Hopf algebra structure has the following form:
$$\d_g\cdot\d_h=\d_{gh},\quad\D (\d_g)=\d_g\ot\d_g,\quad S(\d_g)=\d_{g^{-1}},
\quad\e (\d_g )=1,\d_e=1.$$
The category $Mod_H$ is a category $Mod_G$ of left $G$--modules over $k.$

A braiding element $\s\in H\ot H$, $\s=\sum_{a,b\in G}\s (a,b)\,\d_a\ot\d_b ,$
in the case $k=\Bbb C,$ may be considered as a $\Bbb C$--morphism
$$\hat\s :\Bbb C(G)\lra\Bbb C[G],$$
where $\hat\s (\th_a)=\sum_{b\in G}\s (a,b)\,\d_b.$

There is the following description [L4] of braided elements. Let us fix two
central subgroups $H_1,H_2\subset\Cal Z(G)$ and a group homomorphism
$\phi :\hat H_2\lra H_1,$ where $\hat H_2$ is a dual group for $H_2.$

Then any braiding in the category can be obtained from the following
commutative diagram
$$\CD
\Bbb C(G)@>\hat\s>>\Bbb C[G]\\
@Vr_1VV             @A(r_2)^*AA\\
\Bbb C(H_1)@>\tilde\phi>>\Bbb [H_2]
\endCD
$$
where $r_1:\Bbb C(G)\lra\Bbb C(H_1)$ is the restriction map, and
$(r_2)^*:\Bbb C[H_2]\lra\Bbb C[G]$ is the adjoint of $r_2,$ and $\tilde\phi$
is given by the composition
$$\tilde\phi :\Bbb C(H_1)\overset{\phi^*}\to\lra\Bbb C(\hat H_2)\overset
{\Cal F}\to\lra\Bbb C[H_2],$$
where $\Cal F$ is the Fourier transform.

(3) The same description of the braiding elements holds for any compact
Lie group $G,$ [L4].

(4) Let $G$ be a finite group. To introduce a braiding in the category of $G$--
graded modules we need some additional structures.

Consider for example $G$--graded modules $X=\sum_{g\in G}X_g$ equipped with
$G$--action, such that $h(X_g)\subset X_{ghg^{-1}}.$

In this case our Hopf algebra $H$ is a smash product $k(G)\# k[G],$ i.e. a Hopf
algebra generated by products $\th_g\,\d_h$ with the following relations:
$\d_h\,\th_g=\th_{hgh^{-1}}\,\d_h.$

Define a braiding in this category as follows:
$$\s_{X,Y}(x_g\ot y_h)=g(y_h)\ot x_g,\tag1$$
where $X=\sum_{g\in G}X_g,\quad Y=\sum_{h\in G}Y_h,$ and
$x_g\in X_g,y_h\in Y_h.$

It is easy to check that $\s$ is a braiding.

One can deform the braiding by some function $s:G\times G\lra U(k):$
$$\s_{X,Y}(x_g\ot y_h)=s(g,h)\,g(y_h)\ot x_g.\tag2$$
Then $\s$ is a braiding in the category if the following conditions hold
$$s(aga^{-1},aha^{-1})=s(g,h),\tag3$$
$$s(f,hg)=s(f,h)\,s(f,g),\tag4$$
and
$$s(fh,g)=s(f,hgh^{-1})s(h,g),\tag5$$
for all $a,f,gh\in G.$

We say that a normalized function $s,\quad s(e,g)=s(g,e)=1,$ is a {\it colour}
on the group if conditions (3,4,5) hold.

Note that as a rule braidings (1) and (2) are not symmetries.

The braiding element $\s$ in this case has the following form:
$$\s=\sum_{g,h\in G}s(g,h)\d_g\th_h\ot\th_g.$$
\endexample


{\bf 1.4.} Let $A$ be an algebra in the category and let  $X$ be $A-A$
bimodule in the category.

Denote by $\mu :A\ot A\lra A,\quad\mu (a\ot b)=a\cdot b,$ a multiplication in
the algebra and by $\mu^r:X\ot A\lra X,\quad\mu^r(x\ot a)=xa$ the right
multiplication and by $\mu^l:A\ot X\lra X,\quad\mu^l(a\ot x)=ax,$ the left
multiplication in the bimodule.

By using a braiding $\s$ we introduce new multiplications:
$$\mu_\s =\mu\circ\s_{A,A},\quad\mu^l_\s=\mu^r\circ\s_{A,X},\quad\mu^r_\s =
\mu^l\circ\s_{X,A},$$
and denote the new multiplications by ${}*{}$:
$$a*b=\mu_\s (a\ot b),\quad a*x=\mu^l_\s (a\ot x),\quad x*a=\mu^r_\s
(x\ot a),$$
where $a,b\in A,x\in X.$

In terms of the braiding element $\s$ we have
$$a*b=\sum\s '(b)\s ''(a),\quad a*x=\sum\s '(x)\s ''(a),\quad x*a=\sum\s '(a)
\s ''(x).$$
\proclaim{Proposition}
\roster
\item The pair $(A,\mu_\s )$ determines an algebra structure in the category.
\item The triple $(X,\mu^l_\s ,\mu^r_\s )$ determines $(A,\mu_\s ) -
(A,\mu_\s )$  bimodule structure in the category.
\endroster
\endproclaim
\demo{Proof} We show, for example, that $\mu^l_\s$ determines a left $(A,
\mu_\s )$--structure and that the multiplications $\mu^r_\s$ and $\mu^l_\s$
commute.

For the first one we have
$$\align
a*(b*x)&=\sum\s '(b*x)\,\s ''(a)=\sum\s_{(1)}'\tilde\s '(x)\,\s_{(2)}\tilde
\s ''(b)\,\s ''(a)\\
&\overset\text{hexagon}\to =\mu^r\circ (id\ot\mu )(\s_{23}\s_{13}\s_{12})
(x\ot b\ot a)\\
\text{and}\qquad&\\
(a*b)*x&=\sum\s '(x)\,\s ''(a*b)=\sum\s '(x)\,\s_{(1)}''\tilde\s
'(b)\,\s_{(2)}''\tilde
\s ''(a)\\
&\overset\text{hexagon}\to =\mu^r(id\ot\mu )(\s_{12}\s_{13}\s_{23})(x\ot b\ot
a).\\
\endalign$$
Hence, the Yang--Baxter equation implies the left $(A,\mu_\s )$-- module
structure.

Comparing terms $(a*x)*b$ and $a*(x*b)$  we get
$$\align
(a*x)*b&=\sum\s '(b)\,\s ''(a*x)=\sum\s '(b)\,\s_{(1)}''\tilde\s
'(x)\,\s_{(2)}''
\tilde\s ''(a)\\
&\overset\text{hexagon}\to =\mu^l\circ (id\ot\mu^r)(\s_{12}\s_{13}\s_{23})
(b\ot x\ot a),\\
\text{and}\qquad&\\
a*(x*b)&=\sum\s '(x*b)\,\s ''(a)=\sum\s '_{(1)}\tilde\s '(b)\,\s_{(2)}'\tilde\s
''(x)\,\s ''(a)\\
&\overset\text{hexagon}\to =\mu^l\circ (id\ot\mu^r)(\s_{23}\s_{13}\s_{12}
(b\ot x\ot a)).\\
\endalign$$
Therefore, $(A,\mu_\s )-(A,\mu_\s )$ bimodule structure follows from
Yang--Baxter equation too.
\qed\enddemo

An algebra $A$ in the category $\C$ is called a $\s$--{\it commutative algebra}
if $\mu_\s =\mu,$ or equivalently, if
$$a\cdot b= a*b=\sum\s '(b)\cdot\s ''(a),$$
for all $a,b\in A.$

An $A-A$ bimodule $X$ in the category $\C$ is called $\s$--{\it symmetric} if
$\mu^r_\s =\mu^r$ and $\mu^l_\s=\mu^l,$ or in terms of braiding element $\s$
if
$$x\cdot a=x*a=\sum\s '(a)\cdot\s ''(x),\quad a\cdot x=a*x=\sum\s '(x)\cdot
\s ''(a),$$
for all $a\in A,x\in X.$

Define a $\s$--{\it symmetric part} $X_\s$ of any $A-A$ bimodule $X$ in the
category as follows:
$$X_\s =\bigl\{ x\in X\bigm\vert a\cdot x=a*x,\quad x\cdot a= x*a,\quad
\forall a\in A\bigr\}.$$
\proclaim{Theorem}Let A be a $\s$--commutative algebra in the category $\C$
 and let $X$ be an $A-A$ bimodule. Then $X_\s$ is a $\s$--symmetric $A-A$
bimodule.
\endproclaim
\demo{Proof} It is enough to show that $ax\in X_\s,\/\text{and}\quad xa\in X,$
if $x\in X_\s,a\in A.$

Let us prove, for example, the first inclusion.

For any elements $a,b\in A,x\in X_\s$ one has
$$b*(ax)=b*(a*x)=(b*a)*x=(ba)*x=(ba)x=b(ax).$$
Hence, $ax\in X_\s .$
\qed\enddemo
\example{Examples}

(1) Let $G$ be an Abelian group. Consider the category of $G$--graded modules
with a braiding $\s$ is given by the group bihomomorphism
$\s :G\times G\lra U(k),$ (see ex.1.3.(1)).

An algebra in this category is a $G$--graded algebra $A=\sum_{g\in G}A_g,\quad
A_g\,A_h\subset A_{gh}.$ The algebra is a $\s$--commutative if and only if the
following relations hold
$$a_g\,a_h=\s (g,h)\,a_h\,a_g,$$
for all $a_g\in A_g,\quad a_h\in A_h.$

(2) One can reformulate the FRT-construction [FRT] of function algebras on
quantum groups in the case of $G$--graded modules in the following way.

Let $X=\sum_{g\in G}X_g$ be a $G$--graded module and $X^*=\sum_{g\in g}X^*_g$
be the dual:$X^*_g=(X_{g^{-1}})^*.$ Consider a new module $Y=X\ot X^*.$ This
module is generated by elements $y_{g,h}=x_g\ot y^*_h,$ of degree $gh^{-1}$
and the braiding takes the form:
$$\s_{X,Y}:y_{a,b}\ot y_{c,d}\mapsto\s (ab^{-1},cd^{-1})\,y_{c,d}\ot y_{a,b}.$$

The $\s$--symmetric algebra generated by elements of $y_{a,b}$ is a factor
of the tensor algebra $T(Y)$ by two-sided ideal generated by $Im(\s -1).$

Note that this algebra is not trivial for any braiding $\s .$

(3) For the case of the category of $G$-graded $G$--modules (see ex.1.3.(3))
an algebra in the category is a $G$--graded algebra $A=\sum_{g\in G}A_g$
equipped with $G$--action: $g(A_h)\subset A_{ghg^{-1}}.$

The condition of $\s$--commutativity for the given colour $s$ takes the form:
$$a_g\cdot a_h=s(g,h)\,g(a_h)\cdot a_g.$$

(4) Consider the group algebra $k[G]$ equipped with a $G$ action:
$$g(\d_h)=\d_{ghg^{-1}},$$
as an algebra in the category of $G$--graded $G$--modules.

The algebra will be $\s$--commutative with respest to braiding (1):
$$\d_g\cdot\d_h=g(\d_h)\cdot\d_g.$$

(5) Crossed products.

Let $\om :G\times G\lra U(k)$ be a multiplicative $G$--invariant
$(\om (aga^{-1},aha^{-1})=\om (g,h),\quad\forall a,g,h\in G)$ 2-cocycle on the
group.

The crossed product $k_{\om}[G]$ coincides with $k[G]$ as a $G$--graded $G$--
module but has a new multiplication:
$$\d_g*\d_h=\om (g,h)\d_{gh}.$$
In this case the function
$$s(g,h)=\om (g,h)\cdot{\om (h,g)}^{-1}$$
determines a colour on the group, and $k_{\om}[G]$ is a $\s$--commutative
algebra.

(6) Let $G=\Bbb Z^n,\,k=\Bbb C,$ and let $\Theta =\Vert\th_{ij}\Vert
\in Mat_n(\Bbb R)$ be a matrix such that $\th_{ij}=0,$ if $i\le j.$ Taking the
twisting 2-cocycle $\om$ in the form
$$\om (x,y)=exp (\pi\imath\langle\Theta x,y\rangle ), $$
where $x=(x_1,...,x_n),y=(y_1,...,y_n)\in\Bbb Z^n,$ and $\langle x,y\rangle =
\sum x_i y_i,$  we get a $\s$--commutative algebra $\Bbb C_{\om}[\Bbb Z^n]$,
which is called a quantum torus.
\endexample
{\bf 1.5.} Here we apply the above procedure to modules of internal
homomorphisms
and left modules.

Let $P$ and $Q$ be left $A$--modules. Then the module of internal homomorphisms
$Hom(P,Q)$ is endowed with an $A-A$ bimodule structure with respect to left
and right multiplication
$$l_a(f)(p)=af(p),\quad r_a(f)(p)=f(ap).$$
Denote by $Hom_\s (P,Q)\subset Hom(P,Q)$ the $\s$--symmetric part of the
bimodule.

Elements of $Hom_\s (P,Q)$ are called $\s$--{\it homomorphisms.}

Therefore, an internal homomorphism $f:P\lra Q$ is a $\s$--homomorphism if and
only if the following relations hold:
$$f(a\cdot p)=\sum\s '(a)\cdot\s ''(f)(p),\tag1$$
and
$$a\cdot f(p)=\sum\s '(f)(\s ''(a)\cdot p),\tag2$$
for all $a\in A,p\in P.$

Remark that if $f:P\lra Q$ is simultaneously a morphism in the category and a
$\s$--homomorphism, then $h(f)=\e (h)\,f,$ for all $h\in H,$ and
therefore the above conditions are reduced to the following known one:
$f(a\,p)=a\,f(p).$

\proclaim{Theorem} Let $A$ be a $\s$--commutative algebra. Then for any left
$A$--module $P$ with left multiplication $\mu^l:A\ot P\lra P,$  right
multiplication
$\mu^r=\mu^l\circ\s_{P,A}:P\ot A\lra P$ determines $A-A$ bimodule structure.
\endproclaim
\demo{Proof} We should show that the right and the left structures
commute.

One has
$$\align
(ap)b&=\sum\s '(b)\s ''(ap)=\sum\s '(b)\s ''_{(1)}(a)\s_{(2)}''(p)=\\
&\sum\s '\tilde\s '(b)\s ''(a)\tilde\s ''(p)=\sum a\tilde\s '(b)\tilde\s
''(p)=a(bp)\\
\endalign$$

\qed\enddemo

Let $P$ be a left $A$--module over a $\s$--commutative algebra $A.$ Consider
$P$ as $A-A$ bimodule. We introduce a bimodule $P_\s$ which is the $\s$--
symmetric part of the bimodule $P$.

We have the following direct description of the bimodule:
$$P_\s =\bigl\{ p\in P\bigm\vert (\mu^l-\mu^l\circ\s_{P,A}\circ\s_{A,P})
(a\ot p)=0,\quad\forall a\in A\bigr\}.$$
In terms of the braiding element we get
$$p\in P_\s\Leftrightarrow ap=\sum\s '(p)\cdot\s ''(a)\overset{\text{def}\mu^r}
\to {=}\sum\tilde\s '\s ''(a)\cdot\tilde\s ''\s '(p),$$
or, if we set $\g=\s\cdot\tau (\s )=\sum\g '\ot\g '',$ where
$\g '=\sum\tilde\s '\cdot\s '',$ and $\g ''=\sum\tilde\s ''\cdot\s ',$ then
$$P_\s=\bigl\{ p\in P\bigm\vert a\cdot p=\sum\g '(a)\cdot\g ''(p)\bigr\}.$$

The $A$-submodule $P_\s\subset P$ will be called the $\s$--
{\it symmetric part} of the left $A$--module $P$.

A left $A$--module $P$ is called $\s$--{\it symmetric} if $P_\s =P.$

{\bf 1.6.} The following theorem describes the relation between two
natural $\s$--symmetric bimodules associated to a given left module.
\proclaim{Theorem} Let $A$ be a $\s$--commutative algebra and let $P$ be a
left $A$--module in the category $\C$. Then there is an isomorphism between
$Hom_\s(A,P)$ and $P_\s$ is given by the formula
$$f\in Hom_\s(A,P)\mapsto f(1)\in P_\s.$$
\endproclaim
\demo{Proof} Let $f\in Hom_\s (A,P)$ then for any elements $a,b\in A$ we have
$$f(ab)=\sum\s '(a)\cdot\s ''(f)(b).$$
Hence, if we set $b=1,\quad p=f(1),$ we get
$$\align
f(a)&=\sum\s '(a)\cdot\s ''(f)(1)=\sum\s'\s_{(1)}''(f(S(\s_{(2)})(1)))\\
&=\sum\s '(a)\cdot\s_{(1)}''(f(\e S(\s_{(2)}))=\sum\s '(a)\cdot (\s_{(1)}''\e
(\s_{(2)}))p\\
&=\sum\s '(a)\cdot\s ''(p).\\
\endalign$$
{}From condition 1.5.(2) one gets
$$a\cdot f(b)=\sum\s '(f)(\s ''(a)b).$$
Hence,
$$\align
&a\cdot p=\sum\s_{(1)}f(S(\s_{(2)}')\s ''(a))=\sum\s_{(1)}(\tilde\s
'S(\s_{(2)}')
\s ''(a)\cdot\tilde\s ''(p)=\\
&\sum(\s_{(1)}'\tilde\s 'S(\s_{(3)}')\s ''(a))\cdot (\s_{(2)}'\tilde\s ''(p))
\overset{1.3.(3)}\to{=}\sum(\tilde\s '\s_{(2)}'S(\s_{(3)}')\s ''(a)\cdot
(\tilde
\s ''\s_{(1)}'(p))=\\
&\sum(\tilde\s '\e (\s_{(2)}')\s ''(a))\cdot (\tilde\s ''\s_{(1)}'(p))=
\sum(\tilde\s '\s ''(a))\cdot (\tilde\s ''\s '(p)),\\
\endalign$$
where $\s =\sum\s '\ot\s ''=\sum\tilde\s '\ot\tilde\s ''.$

Therefore, $p\in P_\s .$

Similar calculations show that for any element $p\in P_\s$ the formula
$$f_p(a)=\sum\s '(a)\cdot\s ''(p)\tag1$$
determines a $\s$--homomorphism $f_p\in Hom_\s(A,P).$
\qed\enddemo

{\bf 1.7.} Let $A$ be a $\s$--commutative algebra and $X$  an $A-A$ bimodule.

Consider a quotient bimodule $X/X_\s$ and define a bimodule $X_\s^{(1)}\subset
 X$ as the inverse image of the bimodule $(X/X_\s )_\s\subset X/X_\s$ with
respect to the natural projection $X\lra X/X_\s .$

Thus we get an embedding $X_\s\subset X_\s^{(1)}$ and $X_\s^{(1)}/X_\s$ is a
$\s$--symmetric bimodule by construction.

Proceeding in this way, we obtain a filtration of the bimodule $X$ by bimodules
$X_\s^{(i)},\quad i=-1,0,1,...:$
$$0=X_\s^{(-1)}\subset X_\s^{(0)}=X_\s\subset X_\s^{(1)}\subset\cdots\subset
 X_\s^{(i)}\subset X_\s^{(i+1)}\subset\cdots\subset X_\s^{(*)}\subset X$$
where, by definition, $X_\s^{(i+1)}\subset X$ is the inverse image of
$(X/X_\s^{(i)})_\s$ with respect to the projection $X\lra X/X_\s^{(i)}.$

Note that all the quotients $X_\s^{(i)}/X_\s^{(i-1)}$ are $\s$--symmetric
modules by the construction.

We call the bimodule $X_\s^{(*)}=\bigcup X_\s^{(i)}$ a {\it differential
approximation} of the $A-A$ bimodule $X.$

To produce more concrete description of the differential approximation,
we define two types of morphisms:
$$\d^l_a(x)=a\cdot x-a*x,$$
and
$$\d^r_a(x)=x\cdot a-x*a.$$
Then, by definition, we have
$$X_\s=X_\s^{(0)}=\bigl\{ x\in X\bigm\vert\d^l_a(x)=\d^r_a(x)=0,\forall a
\in A\bigr\} ,$$
and
$$X_\s^{(i)}=\bigl\{ x\in X\bigm\vert\d^l_a(x)\in X_\s^{(i-1)},\d_\a^r(x)
\in X_\s^{(i-1)},\forall a\in A\bigl\} .$$
\example{Examples}

(1) In the category of $G$--graded modules over commutative group $G,$ an
$A-A$ bimodule $X=\sum_{g\in G}X_g$  is a $G$--graded bimodule such that
$$A_h\cdot X_g\subset X_{hg},\quad X_g\cdot A_h\subset X_{gh}.$$
The bimodule is $\s$--symmetric if and only if
$$a_g\cdot x_h=\s (g,h)\,x_h\cdot a_g,\quad\text{and}\quad x_h\cdot a_g
=\s (h,g)\,a_g\cdot x_h,$$
for all $x_h\in X_h,\quad a_g\in A_g,\quad g,h\in G.$

The $\s$--symmetric part of any $A-A$ bimodule $X$ is $X_{\s}=\sum_{g\in G}
(X_{\s})_g,$ where
$$(X_{\s})_g=\{ x_g\in X_g\vert a_h x_g=\s (h,g)\,x_g a_h,x_g a_h=\s (g,h)\,
a_h x_g,\forall a_h\in A_h,h\in G\} .$$
The $\d$--morphisms have the following form
$$\align
&\d^l_{a_h}(x_g)=a_h\cdot x_g-\s (h,g)\,x_g\cdot a_h,\\
&\d^r_{a_h}(x_g)=x_g\cdot a_h-\s (g,h)\,a_h\cdot x_g.\\
\endalign
$$
Then
$$(X_{\s})_g=\{ x_g\in X_g\vert\d^l_{a_h}(x_g)=\d^r_{a_h}(x_g)=0,\quad\forall
 a_h\in A_h,h\in G\} ,$$
and
$X^{(i)}_\s =\sum_{g\in G}(X^{(i)}_\s )_g,$ where
$$(X^{(i)}_\s )_g=\{ x_g\in X_g\vert\d^l_{a_h}(x_g)\in (X^{(i-1)}_\s )_{(hg)}
,\d^r_{a_h}(x_g)\in (X^{(i-1)}_\s )_{(gh)},\forall a_h\in A_h\} .$$

(2) In the category of $G$--graded $G$--modules (see ex.1.3.(4)) with braiding
$\s$ given by colour $s$ we have, accordingly,
$$\align
&\d^l_{a_h}(x_g)=a_h\cdot x_g - s(h,g)h(x_g)\cdot a_h,\\
&\d^r_{a_h}(x_g)=x_g\cdot a_h -s(g,h)g(a_h)\cdot x_g,\\
\endalign
$$
and the same description of $X^{(i)}_\s .$
\endexample
{\bf 1.8.} Applying the above procedure to bimodules of internal homomorphisms
$X=Hom(P,Q),$ we obtain modules of {\it braided differential operators}:
$$Diff^\s_i(P,Q)=(Hom(P,Q))_\s^{(i)}$$
in the category $\C$.

Keeping in the mind the definition of differential approximations, we build up
the modules of braided differential operators over quasitriangular Hopf
algebra in a direct way.

To do this, we introduce two types of morphisms in $Hom(P,Q)$:
$$\d^l_a(f)(p)=a\cdot f(p)-\sum\s '(f)(\s ''(a)\cdot p),$$
and
$$\d^r_a(f)(p)=f(a\cdot p)-\sum\s '(a)\cdot\s ''(f)(p),$$
for all $a\in A,p\in P,f\in Hom(P,Q).$

Then
$$Diff^\s_0(P,Q)=Hom_\s(P,Q)=\bigl\{ f\in Hom(P,Q)\bigm\vert\d^r_a(f)=\d^l_a
(f)=0,\forall a\in A\bigr\} ,$$
and
$$Diff^\s_i(P,Q)=\bigl\{ f\in Hom(P,Q)\bigm\vert\d^r_a(f),\d^l_a(f)\in
Diff^\s_{(i-1)}(P,Q),\forall a\in A\bigr\}.$$
\example{Examples}

(1) In the category of $G$--graded modules we have
$$\align
&\d^l_{a_h}(f)(x)=a_h\,f(x) - \s (h,g)\,f(a_h x),\\
&\d^r_{a_h}(f)(x)=f(a_hx) - \s (g,h)\,a_h f(x),\\
\endalign
$$
if $f\in Hom(X,Y)$ is a homomorphism of degree $g.$

In the case $\d^l_{a_h}(f)$ and $\d^r_{a_h}(f)$ are homomorphisms of degree
$hg.$

(2) In the category of $G$--graded $G$--modules we have
$$\align
&\d^l_{a_h}(f)(x)=a_h\,f(x)-s(h,g)h(f)(a_hx),\\
&\d^r_{a_h}(f)(x)=f(a_h x)-s(g,h)g(a_h)\,f(x),\\
\endalign$$
where the homomorphism $f$ has degree $g$ and $h(f)(x)=h(f(h^{-1}x)).$

(3) The quantum hyperplane is given by the following data: $k=\Bbb C,\quad
G=\Bbb Z^n,$ and the twisted 2-cocycle
$$\om (\bar a,\bar b)=q^{\langle\Theta\bar a,\bar b\rangle},$$
where $\Theta$ is a skew symmetric $n\times n$ matrix, $q\in\Bbb C^*,\quad
\bar a,\bar b\in\Bbb Z^n.$

Let $A$ be a $\s$--commutative algebra in the category of $G$--graded modules.
Assume that $A$ is generated by elements $x_1,...,x_n$ and the relations
$$x_i\cdot x_j=\om_{ij}\,x_j\cdot x_i,$$
where $\om_{ij}$ are matrix elements of $\om .$

Then the algebra of differential operators $Diff^\s_*(A,A)$ is a $\Bbb Z^n$--
graded algebra generated by elements $x_i$ of degree $1_i=(0,...,0,\underset i
\to 1,0,...,0)\in\Bbb Z^n$ and operators $\partial_i$ of degree $-1_i$ and the
following relations
$$\align
&\partial_i\cdot x_j - {\om_{ij}}^{-1}\, x_j\cdot\partial_i=\d_{ij},\\
&\partial_i\cdot\partial_j -\om_{ij}\,\partial_j\cdot\partial_i=0,\\
&x_i\cdot x_j -\om_{ij}\,x_j\cdot x_i=0.\\
\endalign$$

\endexample
{\bf 1.9.} In this section we show that the composition of braided
differential operators is a braided differential operator.

We start with the following lemmas.
\proclaim{Lemma 1} Let $A$ be a $\s$--commutative algebra and $P,Q,R$
left $A$--modules. Then for any internal homomorphisms $f\in Hom(Q,R),g\in
Hom(P,Q)$ and $h\in H$ we have:
$$h(f\circ g)=\sum_h h_{(1)}(f)\circ h_{(2)}(g).$$
\endproclaim
\demo{Proof} From the definition of $H$--action on the modules of internal
homomorphisms we have
$$h(f\circ g)(p)=\sum_h h_{(1)}((f\circ g)(S(h_{(2)})p).$$
On the other side we have
$$\align
&\sum_h h_{(1)}(f)\circ h_{(2)}(g)(p)=\sum_h
h_{(1)}(f)(h_{(2)}(g(S(h_{(3)}p)))\\
&=\sum_h h_{(1)}(f(S(h_{(2)}h_{(3)}(g(S(h_{(4)}p))))=\sum_h h_{(1)}
\e (h_{(2)})(f(g(S(h_{(3)}p)))\\
&=\sum_h h_{(1)}((f\circ g)(S(h_{(2)})p)).\\
\endalign$$
\qed\enddemo
\proclaim{Lemma 2} The following formulae hold
$$\left\{\aligned
&\d^r_a(f\circ g)=f\circ\d^r_a(g)+\sum\d^r_{\s '(a)}(f)\circ\s ''(g),\\
&\d^l_a(f\circ g)=\d^l_a(f)\circ g+\sum\s '(f)\circ\d^l_{\s ''(a)}(g),\\
\endaligned\right.$$
for all $a\in A,f\in Hom(Q,R),g\in Hom(P,Q).$
\endproclaim
\demo{Proof} To prove the first formula it is enough to compare
$$\align
&\d^r_a(f\circ g)(p)=(f\circ g)(ap)-\sum\s '(a)(\s ''_{(1)}(f)\circ
\s ''_{(2)}(g))(p)\\
&\overset\text{hexagon}\to =(f\circ g)(ap)-\sum\s '\tilde\s '(a)\cdot\s ''(f)
\circ\tilde\s ''(g)(p),\\
\endalign$$
and
$$\align
&f\circ\d^r_a(g)(p)=(f\circ g)(ap)-\sum f(\s '(a)\s ''(g)(p))\\
&=(f\circ g)(ap)-\sum\d^r_{\s '(a)}(f)(\s ''(g)(p))-\sum\tilde\s '(\s '(a))
\tilde\s ''(f)(\s ''(g)(p)).
\endalign$$
Similarly, for the second formula we get
$$\align
&\d^l_a(f\circ g)(p)=a(f\circ g)(p)-\sum\s '(f\circ g)(\s ''(a)p)\\
&=a(f\circ g)(p)-\sum(\s '_{(1)}(f)\circ\s '_{(2)}(g))(\s ''(a)p)\\
&\overset\text{hexagon}\to =a(f\circ g)(p)-\sum(\tilde\s '(f)\circ\s '(g))
(\s ''\tilde\s ''(a)p),\\
\endalign$$
and
$$\align
&a(f\circ g)(p)=(\d^l_a(f)\circ g)(p)-\sum\s '(f)(\s ''(a)g(p))=\\
&(\d^l_a(f)\circ g)(p)-\sum(\s '(f)\circ\d^l_{\s ''(a)}(g))(p)+\sum\s '(f)
\circ\tilde\s '(g)(\tilde\s ''\s ''(a)(p).\\
\endalign$$
\qed\enddemo
\proclaim{Lemma} Maps $\d^l:A\ot Hom(P,Q)\lra Hom(P,Q)$ and $\d^r:Hom(P,Q)\ot A
\lra Hom(P,Q)$ are $H$--morphisms:
$$h(\d^l_a(f))=\sum_h\d^l_{h_{(1)}(a)}(h_{(2)}(f)),$$
and
$$h(\d^r_a(f))=\sum_h\d^r_{h_{(2)}(a)}(h_{(1)}(f)),$$
for all $a\in A,\quad h\in H,\quad f\in Hom(P,Q).$
\endproclaim
\proclaim{Theorem}Let $A$ be a $\s$--commutative algebra and let $P,Q,R$ be
left $A$--modules in the category $\C.$

Then
\roster
\item $f\in Diff^\s_i(Q,R),g\in Diff^\s_j(P,Q)\Longrightarrow f\circ g\in
Diff^\s_{i+j}(P,R).$
\item $f\in Diff^\s_i(A,A),g\in Diff_j^\s(A,A)\Longrightarrow [f,g]_\s\in
Diff^\s_{i+j-1}(A,A),$
\endroster
where $[f,g]_\s=f\circ g-\sum\s '(g)\circ\s ''(f)$ is a $\s$--commutator of
internal homomorphisms.
\endproclaim
\demo{Proof} The first part of the theorem is follows from the lemmas. The
second part is a consequence of the definition of braided differential
operators and part (1) of the theorem.
\qed\enddemo

\head{\bf 2.Braided Calculus}\endhead

In this chapter we introduce braided derivations as special 1-st order braided
differential operators and consider braided differential forms as a
representative object for the functor of braided derivations. We construct
an algebra of braided differential forms as a new $\hat\s$--commutative
algebra equipped with a universal braided derivation $d$ such that $d^2=0.$
We show that the main facts of the usual calculus can be translated in the
case of arbitrary braidings.

{\bf 2.1.} Let $A$ be a $\s$--commutative algebra and $P$ be a left $A$--module
in the category $\C$.

We define modules of {\it braided derivations} as follows
$$D(P)=\{\left. f\in Diff^\s_1(A,P)\right\vert f(1)=0\}.$$

Elements of $D(P)$ will be called {\it braided} or $\s$--{\it derivations}
of the algebra $A$ with values in the module $P$.

We now produce a description of $\s$--derivation in terms of generalized
 (or braided) Leibniz rule.

Let $f\in D(P)$ and $a\in A.$ Then $\d^l_a(f),\d^r_a(f)\in Hom_\s(A,P)
=P_\s$ by definition of $\s$--derivation.

Therefore, by 1.6.(1), we have
$$\left\{\aligned
&\d^r_a(f)(b)=\sum\s '(b)\cdot\s ''(p_r),\\
&\d^l_a(f)(b)=\sum\s '(b)\cdot\s ''(p_l),\\
\endaligned\right.$$
for some elements $p_r,p_l\in P_\s.$

We have
$$\left\{\aligned
&p_l=\d^l_a(f)(1)=a\cdot f(1)-\sum\s '(f)(\s ''(a)),\\
&p_r=\d^r_a(f)(1)=f(a)-\sum\s '(a)\cdot\s ''(f)(1).\\
\endaligned\right.$$
Since, $f(1)=0,$ and $h(f)(1)=h_{(1)}\cdot f(S(h_{(2)}(1))=h_{(1)}\e (h_{(2)})
f(1)=0,$ for all $h\in H,$ we get
$$\left\{\aligned
&p_l=-\sum\s '(f)(\s ''(a)),\\
&p_r=f(a).\\
\endaligned\right.$$
Therefore, conditions $p_r\in P_\s,p_l\in P_\s$ mean that $f:A\lra P_\s\subset
P,$
and from formulae (1) and (2) we obtain the following form of the {\it braided
 Leibniz rule}
$$
\left\{
\aligned
&f(a\cdot b)=f(a)\cdot b+\sum\s '(a)\cdot\s ''(f)(b),\\
&\sum\s '(f)(\s ''(a)\cdot b)=a\cdot f(b)+\sum\s '(f)(\s ''(a))\cdot b.\\
\endaligned
\right.\tag1$$
Summarizing, we obtain the following description of braided derivations.
\proclaim{Proposition} An internal homomorphism $f:A\lra P$ is a $\s$--
derivation if and only if
\roster
\item $f:A\lra P_\s\subset P,$ and
\item braided Leibniz rule (1) holds.
\endroster
\endproclaim
\remark{Remark} Let $f:A\lra P$ be a braided derivation and a morphism in the
category. Then $h(f)=\e (h)\cdot f,$ for all $h\in H,$ and the braided
Leibniz rule takes the usual form:
$$f(ab)=f(a)b+af(b).$$
\endremark
\example{Examples}

(1) Let $A$ be an algebra in the category of $G$--graded modules. Then an
internal homomorphism $f:A\lra A$ of degree $g\in G$ is a braided derivation
if the following form of the braided Leibniz rules hold:
$$f(a_h\,b)=f(a_h)\,b+\s (g,h)a_h\,f(b),\tag1$$
and
$$\s (h,g)\,f(a_hb)=a_h\,f(b)+\s (h,g)\,f(a_h)\,b,\tag2$$
for all $a_h\in A_h,\quad b\in A.$

Therefore, in the algebra the folowing relations hold:
$$(\s (h,g)\s (g,h)-1)a_h\,f(b)=0.$$

Remark that we need formula (1) only if $\s$ is a symmetry.

(2) The braided Leibniz rules in the category of $G$--graded $G$--modules take
the form:
$$f(a_hb)=f(a_h)\,b+s(g,h)g(a_h)\,f(b)\tag1,$$
and
$$s(h,g)h(f)(a_hb)=a_h\,f(b)+s(h,g)h(f)(a_h)\,b,\tag2$$
where $f$ as above is an internal homomorphism of degree $g\in G$ and $a_h\in
A_h,\quad b\in A.$

Formula (2) implies the following relation
$$a_g\,f(b)=s(h,g)s(g,h)\,(hgh^{-1})(a_h)\,h(f)(b),$$
for all $a_h\in A_h,\quad b\in A.$

(3)Consider the group algebra $k[G]$ as a $\s$-commutative algebra in the
category of $G$--graded $G$--modules.

Any braided derivation $f:k[G]\lra k[G]$ of degree $g\in G$ is determined by
some function $\nu :G\lra k,$ where $f(\d_h)=\nu (h)\,\d_{gh}.$

{}From the braided Leibniz rules we get the following relations on the
function:
$$\left\{\aligned
&\nu (h_1h_2)=\nu (h_1)+\nu (h_2),\\
&\nu (h_1^{-1}h_2h_1)=\nu (h_2),\\
\endaligned\right.$$
for all $h_1,h_2\in G.$

Remark that the second condition is a consequence of the first one.

(4) The $A$--module of derivations of the quantum hyperplane is generated by
operators $\partial_i$ of degree $-1_i$ such that $\partial_i(x_j)=\d_{ij}.$

The Leibniz rule produces the following commutation relations:
$$\partial_i\cdot x_j-\om_{ij}^{-1}x_j\cdot\partial_i=\d_{ij}.$$

\endexample
{\bf 2.2.}Below we collect together the properties of braided derivations.
\proclaim{Theorem}
\roster
\item $D(P)$ is a $\s$--symmetric $A$--module, i.e. $D_\s (P)=D(P),$ or
$$a\cdot f=\sum\g '(a)\cdot\g ''(f),$$
for all $a\in A,\quad f\in D(P).$
\item The module $D(A)$ is closed with respect to the braided commutator
$$[f,g]_\s =f\circ g-\sum\s '(g)\circ\s ''(f),$$
and the commutator is an $H$--invariant:
$$h([f,g]_\s )=\sum [h_{(1)}(f),h_{(2)}(g)]_\s ,$$
for all $f,g\in D(A),\quad h\in H.$
\item Let $f_i\in D(A),\quad i=1,2,3$ be braided derivations of the algebra
$A$ such that
$$f_1\circ f_2=\sum\g '(f_1)\circ\g ''(f_2),\tag1$$
and
$$\sum f_1\circ\s '(f_3)\circ\s ''(f_2)=\sum\g '(f_1)\circ\s '(f_3)\circ\s ''
(\g ''(f_2)).\tag2$$

Then the braided Jacobi identity holds:
$$[f_1,[f_2,f_3]_\s]_\s =[[f_1,f_2]_\s ,f_3]_\s +\sum [\s '(f_1),[\s ''(f_2),
f_3]_\s]_\s .\tag3$$

\endroster
\endproclaim
\demo{Proof}
\roster
\item Using the definition of braided derivations, we get
$$\align
&\sum\g '(a)\g ''(f)(b)=\sum\s '\tilde\s ''(a)\,\s ''\tilde\s '(f)(b)=\\
&\sum\tilde\s '(f)(\tilde\s ''(a)b)-\sum\tilde\s '(f)(\tilde\s ''(a))\,b=\\
&a\cdot f(b).\\
\endalign$$
\item The $\s$-- commutator is a braided differential operator of the order 1,
 and $[f,g]_\s (1)=0.$

Therefore, $[f,g]_\s\in D(A).$

To prove $H$-invariance of the commutator we have:
$$\align
&h([f,g]_\s )=h(f\circ g)-\sum h(\s '(g)\circ\s ''(g))=\\
&\sum_h h_{(1)}(f)\circ h_{(2)}(g)-\sum_h h_{(1)}\s '(g)\circ h_{(2)}\s
''(f)=\\
&\sum_h h_{(1)}(f)\circ h_{(2)}(g)-\sum_h\s 'h_{(2)}(g)\circ\s ''h_{(1)}(f)=\\
&\sum_h[h_{(1)}(f),h_{(2)}(g)]_\s .\\
\endalign$$
\item We have
$$\align
&[f_1,[f_2,f_3]_\s ]_\s = f_1\circ f_2\circ f_3-\sum f_1\circ\s '(f_3)\circ\s
 ''(f_2)-\\
&\sum\s '_{(1)}(f_2)\circ\s '_{(2)}(f_3)\circ\s ''(f_1)+
\sum\tilde\s '\s ''_{(2)}(f_3)\circ\tilde\s ''\s '_{(1)}(f_2)\circ\s ''
(f_1),\\
&[[f_1,f_2]_\s ,f_3]_\s = f_1\circ f_2\circ f_3-\sum\s '(f_2)\circ\s ''(f_1)
\circ f_3-\\
&\sum\s '(f_3)\circ\s ''_{(1)}(f_1)\circ\s ''_{(2)}(f_2)+
\sum\s '(f_3)\circ\tilde\s '\s ''_{(2)}(f_2)\circ\tilde\s ''\s ''_{(1)}
(f_1),\\
\text{and} & {}\\
&\sum[\s '(f_2),[\s ''(f_1),f_3]_\s ]_\s = \sum\s '(f_2)\circ\s ''(f_1)
\circ f_3-\\
&\sum\s '(f_2)\circ\tilde\s '(f_3)\circ\tilde\s ''\s ''(f_1)-
\sum\s '_{(1)}\tilde\s ''(f_1)\circ\s '_{(2)}(f_3)\circ\s ''\tilde\s '(f_2)+\\
&\sum\s '\tilde\s '_{(2)}(f_3)\circ\s ''\tilde\s '_{(1)}\bar\s ''(f_1)\circ
\tilde\s ''\bar\s '(f_2).\\
\endalign $$
\endroster
Comparing coefficients of terms with  $f_if_jf_k,\quad i,j,k=1,2,3$ we see
that they  are equal in the following cases:
\roster
\item $f_2f_3f_1\Longrightarrow$ by the hexagon equations,
\item $f_2f_1f_3\quad\text{and}\quad f_1f_2f_3\Longrightarrow$ are
simply equal,
\item $f_3f_2f_1\Longrightarrow$ by the Yang--Baxter equation.
\endroster
The rest of the Jacobi identity composed of the terms $f_1f_3f_2$ and
$f_3f_1f_2$ is the following:
$$\align
&\sum f_1\circ\s '(f_3)\circ\s ''(f_2)-\sum\s '_{(1)}\tilde\s ''(f_1)\circ
\s '_{(2)}(f_3)\circ\s ''\tilde\s '(f_2)+\\
&\sum\s '\tilde\s '_{(2)}(f_3)\circ\s ''\tilde\s '_{(1)}\bar\s ''(f_1)\circ
\tilde\s ''\bar\s '(f_2)-\sum\s '(f_3)\circ\s ''_{(1)}(f_1)\circ\s ''_{(2)}
(f_2)=\\
&\sum f_1\circ\s '(f_3)\circ\s ''(f_2)-\sum\g '(f_1)\circ\s '(f_3)\circ\s ''
\g ''(f_2)+\\
&\sum\s '(f_3)\circ{\s ''}_{(1)}\g '(f_1)\circ{\s ''}_{(2)}\g ''(f_2)-\sum\s
'(f_3)
\circ{\s ''}_{(1)}(f_1)\circ{\s ''}_{(2)}(f_2).\\
\endalign$$
\qed\enddemo
\proclaim{Corollary 1} The braided Jacobi identity holds if $f_1$ or $f_2$ is
an $H$--invariant braided derivation.
\endproclaim
\demo{Proof}
Suppose, for instance, that $f_1$ is an $H$--invariant derivation. Then
$$\sum\g '(f_1)\circ\g ''(f_2)=\sum\e (\g ')f_1\circ\g ''(f_2)= f_1\circ f_2.$$
In the same way we get condition (2).
\qed\enddemo
\proclaim{Corollary 2} Let $\s$ be a symmetry in the category. Then
\roster
\item The braided Jacobi identity holds for all braided derivations $f_1,f_2,
f_3.$
\item $[f_1,f_2]_\s =-\sum [\s '(f_2),\s ''(f_1)]_\s .$
\endroster
\endproclaim
{\bf 2.3.} In this section we build up the representative object for the
functor
of braided derivations $D:P\lra D(P).$ To do this, we look at $\s$--symmetric
bimodule $\Om^1(A),$ generated by formal elements $a\,db,$ where $a,b\in A,$
with following relations:
\roster
\item $H$--action
$$h(a\,db)=\sum_h h_{(1)}(a)\,dh_{(2)}(b),$$
\item the right $A$--module structure
$$db\cdot a=\sum\s '(a)\,d\s ''(b)$$
and $\s$--symmetric relations
$$a\,db=\sum\g '(a)\,d\g ''(b),$$
where $\g=\s\cdot\tau (\s )=\sum\g '\ot\g '',$ and
\item the usual differential relations
$$d(a+b)=da+db,\qquad d(ab)=da\cdot b+a\cdot db.$$
\endroster
Denote by $d:A\lra\Om^1(A)$ the operator: $d:a\mapsto da.$

The properties above imply that $d$ is a braided derivation.
\proclaim{Theorem} For any braided derivation $f:A\lra P$ there is a $\s$--
homomorphism $\hat f:\Om^1(A)\lra P$ such that
$$f=\hat f\circ d.$$
The $\s$--homomorphism $\hat f$ is uniquely determined, and the
correspondence $f\mapsto\hat f$ establishes an isomorphism in the category
between $D(P)$ and $Hom_\s (A,P_\s ).$
\endproclaim
\demo{Proof} We define $\hat f$ as follows
$$\hat f(a\,db)=\sum\s '(a)\s ''(f)(b),$$
for all $a,b\in A.$

At first we show that $\hat{}$ is a morphism in the category.

One has
$$\align
&h(\hat f)(a\,db)=\sum_h h_{(1)}(\hat f(Sh_{(3)}(a)\,d(Sh_{(2)}b)=h_{(1)}
\sum\s '(Sh_{(3)}(a))\s ''(f)(Sh_{(2)}b)=\\
&\sum h_{(1)}\s '(Sh_{(4)}(a)h_{(2)}(\s ''(f)(Sh_{(3)}b))=\sum h_{(1)}\s '
(Sh_{(3)}(a))(h_{(2)}\s '')(f)(b)=\\
&\sum\s '(h_{(2)}Sh_{(3)}a)(\s ''h_{(1)})(f)(b)=\sum\s '(\e (h_{(2)})a)(\s ''
h_{(1)})(f)(b)=\\
&\sum\s '(a)(\s ''h)(f)(b)=\hat {h(f)}(a\,db).\\
\endalign $$
Now we check the first property of $\s$--homomorphisms.

Let $p=c\,db,$ then
$$\align
&\hat f(ap)=\hat f(ac\,db)=\sum\s '(ac)\s ''(f)(b)=\sum\s_{(1)}'(a)\s_{(2)}'
(c)\s ''(f)(b)\\
&\overset\text{hexagon}\to =\sum\tilde\s '(a)\s '(c)\s ''\tilde\s ''(f)(b)=
\sum\tilde\s '(a)\tilde\s ''(\hat f)(c\,db)=\sum\s '(a)\s ''(\hat f)(p).\\
\endalign $$
For the second property we have
$$\align
&\sum\s '(\hat f)(\s ''(a)p)=\sum\s_{(1)}'(\hat f(S\s_{(4)}\s ''(a)\cdot S
\s_{(3)}'(c)\/S\s_{(2)}'(b)=\\
&\sum\s_{(1)}'(\tilde\s '(S\s_{(4)}'\s ''(a)S\s_{(3)}'(c))\tilde\s ''(f)(S
\s_{(2)}'(b)))=\\
&\sum\s_{(1)}'(f(S\s_{(4)}'\s ''(a)\cdot S\s_{(3)}'(c)\cdot S\s_{(2)}'(b)))-\\
&\sum\s_{(1)}'(f(S\s_{(4)}'\s ''(a)S\s_{(3)}'(c))\cdot S\s_{(2)}'(b))=\\
&\sum\s '(f)(\s ''(a)cb)-\sum\s '(f)(\s ''(a)c)\cdot b=\\
&af(cb)+\sum\s '(f)(\s ''(a))cb-af(c)b-\sum\s '(f)(\s ''(a))cb=\\
&af(cb)-af(c)b=a\sum\s '(c)\s ''(f)(b)=a\hat f(c\,db).\\
\endalign $$
\qed\enddemo
{\bf 2.4.} Starting from this point we will build up an algebra of braided
differential forms over a $\s$--commutative algebra $A.$

The algebra will be a new $\hat\s$--commutative algebra
$\Om^*(A)=\sum_{i\in\Bbb N}\Om^i(A),$
for some new braiding $\hat\s $ equipped with some $H$--invariant braided
derivation of degree 1.

The last conditions dictate some restrictions on the braiding $\hat\s .$

To describe these braidings we look at the category $\G$ of $\Bbb N$
--graded objects over $\C .$

The category has for objects families of objects in $\C$ i.e. $X=\{X_n,n\in
\Bbb N\}.$ and for morphisms $f:X\lra Y$ families $f=\{f_n,n\in\Bbb N\}$ of
morphisms $f_n:X_n\lra Y_n$ in $\C .$

We introduce the usual tensor product in $\G$:
$$(X\ot Y)_n=\sum_i X_i\ot Y_{n-i}.$$

Observe that $\G$ has modules of internal homomorphisms

$Hom(X,Y),$ where
$$Hom_n(X,Y)=\{f=\{f_n\}\vert f_n\in Hom(X_i,Y_{i+n})\forall i,n\in\Bbb N\}.$$
We will call elements of $Hom_n(X,Y)$  {\it internal homomorphisms of
degree $n.$}

Below we identify in the usual way an object $Z$ of the category $\C$ with the
object $(Z_n)$ of the category $\G$, where $Z_0=Z,$ and $Z_n=0$ otherwise.

In a similar way, we identify morphisms and internal homomorphisms in $\C$
with morphisms and internal homomorphisms in $\G .$

\proclaim{Theorem} Any braiding $\hat\s_{X,Y}:X\ot Y\lra Y\ot X$ in the
category $\G$ has the form
$$\hat\s_{X,Y}(x_n\ot y_m)=\hat\s_{n,m}\cdot (y_m\ot x_n),$$
 for some family $\{\s_{n,m},n,m\in\Bbb N\}$ of elements of $H\ot H,$
and where $x_n\in X_n,y_m\in Y_m$.

The family $\{\s_{n,m}\}$ is completely determined by the following data:
\roster
\item a braiding $\s =\hat\s_{0,0}$ in the category $\C$,
\item two invertible central group--like elements $\phi ,\psi\in H$,
\item an invertible element $q\in U(k),$
\item for the data $(\s,\phi ,\psi ,q)$ the braiding $\hat\s$ is given
by the formula
$$\hat\s_{n,m}=(q)^{nm}(\phi^n\ot\psi^m)\cdot\s.\tag1$$
\endroster
\endproclaim
\demo{Proof} Rewriting the hexagon conditions for $\hat\s$ in terms of the
family $\hat\s_{n,m}$ we obtain the following relations :
$$\align
& (id\ot\D )\hat\s_{n+m,k}=(\hat\s_{n,k}\ot 1)(\hat\s_{m,k})_{13},\tag2\\
&(\D\ot id)\hat\s_{n,m+k}=(1\ot\hat\s_{n,k})(\hat\s_{nm})_{13},\tag3\\
&\hat\s_{n,m}\cdot\tau (\D h)=\D (h)\cdot\hat\s_{n,m},\quad\forall h
\in H; n,m\in\Bbb N.\tag4\\
\endalign$$
Now by applying the morphism $id\ot\e\ot id$ to the both sides of formula (2)
we get the following recursive relation:
$$ \hat\s_{n+m,k}=(\phi_{n,k}\ot 1)\hat\s_{m,k},\tag5 $$
where
$$\phi_{n,k}=(id\ot\e )\hat\s_{n,k}.\tag6$$
In the same way we get from (3)

$$\hat\s_{n,m+k}=(1\ot\psi_{n,k})\hat\s_{n,m},\tag7$$
where
$$\psi_{n,k}=(\e\ot id)\hat\s_{n,k}.\tag8$$

Letting $n=1$ in (5), we get $\hat\s_{m+1,k}=(\phi_{1,k}\ot 1)\hat
\s_{m,k},$ and therefore
$$\hat\s_{m,k}=(\phi^m_{1,k}\ot 1)\hat\s_{0,k}.$$
In a similar way, letting $k=1$ in formula (6), we get
$\hat\s_{n,m+1}=(1\ot\psi_{n,1})\hat\s_{n,m},$ and therefore
$$\hat\s_{n,m}=(1\ot\psi^m_{n,1})\hat\s_{n,0}.$$
Let $\phi =\phi_{1,0},\psi =\psi_{0,1},\s =\s_{0,0},$ then
$\hat\s_{n,0}=(\phi^n\ot 1)\s ,$ and $\hat\s_{0,k}=(1\ot\psi^k)\s .$

{}From the relations
$$\phi_{1,k}=(id\ot\e )\hat\s_{1,k}=\left\{\aligned
&(id\ot\e )(\phi_{1,k}\ot\psi^k)\s =\e (\psi)^k\phi_{1,k},\\
&(id\ot\e )(\phi\ot\psi^k_{1,1})\s =\e (\psi_{1,1})^k\phi ,\\
\endaligned\right.$$
we get $\e (\psi )=1,$ and $\phi_{1,k}=\e (\psi_{1,1})^k\phi .$

In a similar way, from the relations
$$\psi_{n,1}=(\e\ot id)\hat\s_{n,1}=\left\{\aligned
&(\e\ot id)(\phi_{1,1}^n\ot\psi )\s =\e (\phi_{1,1})^n\psi ,\\
&(\e\ot id)(\phi^n\ot\psi_{n,1})\s =\e (\phi )^n\psi_{n,1},\\
\endaligned\right.$$
we get $\e (\phi )=1,$ and $\psi_{n,1}=\e (\phi_{1,1})^n\psi .$

Now if we put $q=\e (\psi_{1,1})=\e (\phi_{1,1})\in k,$ we obtain formula (1).
By substituting this formula in relations(2),(3) and (4) we find that $\phi$
and $\psi$ are central group--like elements.
\qed\enddemo
{\bf 2.5.} To motivate the following considerations, we assume that
a $\s$--commutative algebra $A$ is embedded in some $\hat\s$--commutative
algebra $\hat A=\sum_{n\in\Bbb N}A_n,\quad A_0=A,$ with wedge multiplication
$\w .$

Assume also that the algebra $\hat A$ is equipped with a non--trivial $H$--
invariant braided derivation $d$.

The braided Leibniz rule for $H$--invariant derivations of the algebra
takes the form
$$\left \{\aligned
&d(\a_n\w\a_m )=d\a_n\w\a_m +{q^n}\phi (\a_n )\w d(\a_m ),\\
&q^n\,d(\psi(\a_n )\w\a_m )=\a_n\w d(\a_m) +q^n\,d(\psi(\a_n ))\w\a_m,\\
\endaligned\right.$$
where $\a_n\in A_n,\a_m\in A_m.$

Comparing these relations shows that $q^{2n}\phi\psi =1$ on $A_n,\quad\forall
 n\in\Bbb N.$
Therefore, $q^2=1$ and $\phi\psi =1$ on $\hat A.$

We should remark also that $[d,d]_{\hat\s}=(1-q)d^2 ,$ and therefore $d^2$ is
a braided derivation if $(1-q)$ is an invertible element of $k.$
\definition{Definition} The braiding $\hat\s$ given by formula 2.4.(1)
with $q=-1$ and $\phi\psi =1$ will be called a {\it differential prolongation}
of the braiding $\s .$
\enddefinition

{\bf 2.6.} Let us fix a braiding $\s$, group--like element $\phi\in H$ and
let $\hat\s$ be the differential prolongation of $\s .$

Denote by $\Om^1(A,\phi )$ a bimodule generated by formal elements
$a\,d_\phi b$, where $a,b\in A,$ with new relations (cf.2.3.)
\roster
\item $H$--action
$$h(a\,d_\phi b)=\sum_h h_{(1)}(a)\,d_\phi h_{(2)}(b),$$
\item the right $A$--module structure
$$d_\phi b\cdot a=\sum\phi\s '(a)\,d_\phi\s ''(b),$$
\item $\hat\s$--symmetric relations
$$a\,d_\phi b=\sum\phi\g '(a)\,d_\phi\g ''(b),$$
\item and new differential relations
$$d_\phi (a+b)=d_\phi (a)+d_\phi (b),\qquad d_\phi (ab)=d_\phi (a)\cdot b +
\phi (a)\,d_\phi (b).$$
\endroster
Note that $\Om^1(A,1)=\Om^1(A),$ when $\phi =1$ on $A.$

Denote by
$$\Om^*(A,\phi )=\sum_{n\in\Bbb N}\Om^n(A,\phi )$$
the $\hat\s$--commutative algebra generated by $A$ and $\Om^1(A,\phi )$ .

Let $d_\phi :\Om^{n}(A,\phi )\lra\Om^{n+1}(A,\phi ),\quad n\in\Bbb N,$ be the
$\hat\s$--derivation of degree 1, defined by the formula
$$\de (\a\w\be )=\de\a\w\be +(-1)^n\phi^n (\a )\w\de\be ,$$
for all $\a\in\Om^n(A,\phi ),\quad\be\in\Om^*(A,\phi ).$

Then $\de^2=0,$ and for any $\s$--commutative algebra $A$ we get the complex:
$$0\lra A\overset\de\to\lra\Om^1(A,\phi )\overset\de\to\lra\cdots\overset\de
\to\lra\Om^n(A,\phi )\overset\de\to\lra\Om^{n+1}(A,\phi )\overset\de\to\lra
\cdots$$
The cohomology of this complex at the term $\Om^n(A,\phi)$ will be denoted by
$H^n(A,\phi )$ and called {\it braided de Rham cohomology} of the algebra $A.$

Note that the structure of a $\hat\s$--multiplicative algebra in
$\Om^*(A,\phi)$ induces the same structure in the braided cohomology algebra
$$H^*(A,\phi )=\sum_{n\in\Bbb N}H^n(A,\phi ).$$
\example{Examples}

(1) In the category of $G$--graded modules the construction of the algebra of
differential forms over $\s$--commutative algebra $A$ depends on invertible
group-like elements $\phi\in k(G).$

Therefore the construction is determined by the group homomorphisms
$\phi :G\lra U(k).$

For instance, for the trivial group $G=\{ e\}$ we a unique algebra but for the
super-case $G=\Bbb Z_2$ we have two algebras of differential forms.

(2) For the case of quantum hyperplane $k=\Bbb C,\quad G=\Bbb Z^n,$ the
homomorphisms $\phi :\Bbb Z^n\lra\Bbb C^*$ have the form $\phi (a)=z^a,$ for
some complex vector $z=(z_1,...,z_n)\in (\Bbb C^*)^n.$

The algebra of differential forms for the given $\phi$ generaded by the
elements
$x_i$ and $dx_j$ and the relations:
$$x_ix_j=\om_{ij}\,x_jx_i,\quad x_i\,dx_j=z_i\om_{ij}\,dx_j\,x_i,\quad
dx_i\wedge dx_j=-z_iz_j\om_{ij}\,dx_j\wedge dx_i.$$
\endexample
{\bf 2.7.} Let $X$ be a left $A$--module in the category $\C$ and let
$n\in\Bbb N.$ We denote by $X_{(n)}$ a left $\M$--module in the category
$\G $ such that
$(X_{(n)})_n=X,$ and $(X_{(n)})_k=0$ otherwise, with obvious multiplication:
$\om\cdot x=0,$ if $\om\in\M,\quad deg\,\om>1,$ and $a\cdot x=ax,$ if
$x\in X,a\in A.$

As above, we may introduce a right $\M$--module structure in $X_{(n)}$:
$$x\cdot a\overset\text{def}\to =\sum\phi^n\s '(a)\,\s''(x),$$
if $a\in A,x\in X,$  and $x\cdot\om =0,$ if $\om\in\M ,deg\,\om>1.$

The following calculation
$$\align
&\sum\hat\s_{0,n}'(x)\hat\s_{0,n}''(a)=\sum\s '(x)\,\phi^{-n}\s ''(a)\overset
\text{def}\to =\\
&\sum\tilde\s '\s ''(a)\,\tilde\s ''\s '(x)=\sum\g '(a)\,\g ''(x)\\
\endalign$$
shows that $X_{(n)}$ is a $\hat\s$--symmetric $\M$--bimodule if $X$ is a
$\s$--symmetric $A$--module.

Let $f:\M\lra P$ be a $\hat\s$--derivation of degree $k$ with values in
$\hat\s$--symmetric $\M$--module $P.$

We consider the restriction $f_0=f\vert_A:A\lra P_k$ as a $k$--homomorphism of
$\hat\s$--commutative algebra $A=A_{(0)}$ into $\hat\s$--symmetric $\M$--module
 $(P_k)_{(k)}.$

These restrictions may be characterized by a new Leibniz rule.
\definition{Definition} Let $X$ be a $\s$--symmetric left $A$--module. An
internal homomorphism $f:A\lra X$ will be called {\it twisted} (or $\phi$--
{\it twisted}) {\it derivation} of degree $k\in\Bbb Z$, if the following
{\it twisted Leibniz rule} holds:
$$\left\{\aligned
&f(ab)=f(a)\,b+\sum\phi^k\s '(a)\,\s ''(f)(b),\\
&\sum\s '(f)(\phi^{-k}\s ''(a)b)=a\,f(b)+\sum\s '(f)(\phi^{-k}\s ''(a))\,b\\
\endaligned\right.$$
\enddefinition
Denote by $D_{\phi ,k}$ the module of all twisted derivations of degree $k.$
\remark{Remarks}
\roster
\item We have $f_0\in D_{\phi ,k}(P_k),$ for the restriction $f_0.$
\item If $f:A\lra P$ is an $H$--invariant twisted derivation of degree $k,$
then the twisted Leibniz rule takes the form:
$$f(ab)=f(a)\,b+\phi^k(a)f(b).$$
\item The differentials $\dek :A\lra\O1{A,\phi^k}$ are twisted $H$--invariant
derivations of degree $k.$
\endroster
\endremark
\proclaim{Theorem} The morphisms $\dek :A\lra\O1{A,\phi^k}$ are universal
twisted derivations of degree $k$ in the following sense.

Any twisted derivation $f:A\lra X$ of degree $k$ may be represented as the
composition $f=\bar f\circ\dek ,$ where $\bar f:\O1{A,\phi^k}\lra X$ determines
$\hat\s$--homomorphism $(\O1{A,\phi^k})_{(k)}\lra X_{(k)}.$ The
correspondence $f\mapsto\bar f$ establishes an isomorphism
$$D_{\phi ,k}(X)\simeq Hom_{\hat\s ,0}(\O1{A,\phi^k}_{(k)},X_{(k)})\simeq
Hom_\s (\O1{A,\phi^k},X).$$
\endproclaim
{\bf 2.8.} We say that a $\hat\s$--derivation $\M\lra P$ is {\it algebraic},
if $f\vert_A=0.$

Denote by $D_k^{\text{alg}}(P)$ the $\M$--module of all the algebraic
derivations of degree $k\in\Bbb Z,$ and by $D_*^{\text{alg}}(P)=
\sum_{k\in\Bbb Z}D_k^{\text{alg}}(P)$ the graded module of all the algebraic
derivations.

Remark that any algebraic derivation is determined by its restriction to
$\O1{A,\phi}.$ Therefore we have $D_k^{\text{alg}}(P)=0,$ if $k<-1.$

Let $\Om^1_{\text{alg},k}(\M )$ be a representative object for the functor of
algebraic derivations of degree $k$, and let $\pa_k^a:\M\lra\Om^1_{\text{alg},
k}(\M )$ be the universal algebraic derivation.

As before, we may consider $\Om^1_{\text{alg},k}(\M )$  as a
$\hat\s$--symmetric
 module generated by formal elements $\a\w\pa^a_k\be,$ where $\a ,\be\in\M$,
with relations
\roster
\item $H$--action
$$h(\a\w\pa_k^a\be )=\sum_h h_{(1)}(\a )\w\pa_k^ah_{(2)}(\be ),$$
\item the right $\M$--module structure
$$\pa_k^a\be\w\a =\sum\phi^k\s '(\a )\w\pa_k^a\s ''(\be ),$$
and the $\hat\s$--symmetric relations
$$\a\w\pa_k^a\be =\sum\g '(\a )\w\pa_k^a\g ''(\be ),$$
\item algebraic differential relations
$$\pa_k^a(\a +\be )=\pa_k^a\a +\pa_k^a\be,\qquad\pa_k^a(\a\w\be )=\pa_k^a(\a )
\w\be +\phi^k(\a )\w\pa_k^a(\be ),$$
and
$$\pa_k^a(x)=0,$$
for all $x\in A.$
\endroster
Let $P$ be a $\M$--module in the category $\G$. We will denote by $P_{(k)}$ the
shifted module: $(P_{(k)})_n=P_{n+k},\quad n\in\Bbb Z.$
\proclaim{Theorem} Any algebraic derivation $f:\M\lra P,$ of degree $k\ge -1,$
may be represented in a unique way as a composition $f=\bar f\circ\pa_k^a ,$
where
$$\bar f:\Om^1_{\text{alg},k}(\M )\lra P$$
determines $\hat\s$--homomorphism $(\Om_{\text{alg},k}^1(\M ))_{(k)}
\lra P_{(k)}$ and the map $f\mapsto\bar f$ establishes an isomorphism
$$D^{\text{alg}}_k(P)\simeq Hom_{\hat\s ,0}(\Om^1_{\text{alg},k},P)\simeq
Hom_{\hat\s ,0}(\Om^1_{\text{alg},k},P_{(k)}).$$
\endproclaim
{\bf 2.9.} Now we look at the algebra $\M$ as a new $\hat\s$--commutative
algebra and build up the new universal module of braided differential forms
$\O1{\M}$ in the category $\G$ together with new universal $\hat\s$--derivation
$$\pa :\M\lra\O1\M ,$$
of degree 0.

We may consider $\O1\M$ as a $\hat\s$--symmetric $\Bbb Z$--graded module
$$\O1\M =\sum_k{\O1\M}_k,$$
where $\O1\M_k$ is generated by formal elements $\a\w\pa_k\be ,$ where $\a ,\be
\in\M, $ with following relations:
\roster
\item $H$--action
$$h(\a\w\pa_k\be )=\sum_h h_{(1)}(\a )\w\pa_kh_{(2)}(\be ),$$
\item the right $\M$--module structure
$$\pa_k\be\w\a =\sum\phi^k\s '(\a )\w\pa_k\s ''(\be ),$$
and the $\hat\s$--symmetric relations
$$\a\w\pa_k\be =\sum\g '(\a )\w\pa_k\g ''(\be ),$$
\item twisted differential relations
$$\pa_k(\a +\be )=\pa_k\a +\pa_k\be ,\quad\pa_k(\a\w\be )=\pa_k\a\w\be +
\phi^k(\a )\w\pa_k\be.$$
\endroster
Summarizing, we obtain the following
\proclaim{Theorem} The pair $\left(\O1\M ,\pa =\sum\pa_k\right)$ is a
representative object for the functor of graded braided derivations of the
algebra $\M ,$ and
\roster
\item the restriction map $f\in D_k(P)\mapsto f\vert_A\in D_{\phi,k}(P_k)$
defines an embedding
$$0\lra\O1{A,\phi^k}\lra{\O1\M}_k,$$
where $a\,\dek b\mapsto a\,\pa_kb,\quad a,b\in A,$
\item the embedding $D_*^{\text{alg}}(P)\subset D_*(P)$ defines epimorphisms
$${\O1\M}_k\lra\Om^1_{\text{alg},k}(\M )\lra 0,$$
such that $\a\w\pa_k\be\mapsto\a\w\pa_k^a\be ,$
\item the sequence
$$0\lra\O1{A,\phi^k}\lra{\O1\M}_k\lra\Om^1_{\text{alg},k}(\M )\lra 0$$
is exact.
\endroster
\endproclaim
{\bf 2.10.} In this section we describe the module of braided $\hat\s$--
derivations of the algebra braided differential forms $\M .$

We start with an explicit description of algebraic braided derivations. Any
such derivation $f:\M\lra\M $ of degree $k$ is completely determined by the
restriction on $\O1{A,\phi}$  and therefore we obtain an isomorphism:
$$D^{\text{alg}}_k(\M )\simeq Hom_\s (\O1{A,\phi},\O{k+1}{A,\phi}).$$
The image $\imath_\l\in D^{\text{alg}}_k(\M )$ of the element $\l\in Hom_\s
(\O1{A,\phi},\O{k+1}{A,\phi})$ will be called  {\it inner braided derivation.}
  One can define $\imath_\l$ directly  using the braided Leibniz rule:

the derivations
$$\imath_\l :\O{j}{A,\phi}\lra\O{j+k}{A,\phi}$$
are determined by the following relations:
\roster
\item $\imath_\l (a)=0,\quad\forall a\in A=\O0{A,\phi},$
\item $\imath_\l (\om )=\l (\om ),\quad\forall\om\in\O1{A,\phi},$
\item $h(\imath_\l )=\imath_{h(\l )},\quad\forall h\in H,$
\item $\imath_\l (\om_1\w\om_2)=\imath_\l (\om_1)\w\om_2+(-1)^{kj}\sum\phi^k
\s '(\om_1)\w\phi^{-j}\s ''(\om_2),$
\item $\sum\phi^j\imath_{\s '(\l)}(\phi^{-k}\s ''(\om_1)\w\om_2)=\om_1\w
\imath_\l (\om_2)+\sum\phi^j\imath_{\s '(\l )}(\phi^{-k}\s ''(\om_1))\w\om_2,$
\endroster
for all $\om_1\in\O{j}{A,\phi},\om_2\in\O*{A,\phi}.$

Denote by
$$\N{k} =Hom_\s (\O{1}{A,\phi},\O{k+1}{A,\phi})$$
and let
$$\N{*}=\sum_{k\in\Bbb Z}\N{k} .$$
We will call this module  a {\it Nijenhuis algebra} of the $\s$--commutative
algebra $A.$

The discussion above shows that we have isomorphisms
$$\N{k}\simeq {D_k}^{\text{alg}}(\M ).$$
 Modules of braided derivations are closed with
respect to braided commutators.

Therefore we obtain a bilinear structure:
$$[,]_{\hat\s}:\N{k}\times\N{l}\lra\N{k+l},$$
in the $\N{*} .$

Here we define $[\l_1,\l_2 ]_{\hat\s}$ from the following relation:
$$\imath_{[\l_1,\l_2 ]_{\hat\s}}=[\imath_{\l_1},\imath_{\l_2}]_{\hat\s}.$$
The bracket $[,]_{\hat\s}$ will be called {\it braided algebraic Nijenhuis
bracket.}

{}From the definition of the bracket we have:
$$[\l_1,\l_2 ]_{\hat\s}=\imath_{\l_1}(\l_2(\om ))-(-1)^{kl}\sum\imath_
{\phi^k\s '(\l_2)}(\phi^{-l}\s ''(\l_1)),$$
where $\om\in\O{1}{A,\phi}\quad\l_1\in\N{k} ,\l_2\in\N{l} .$

{\bf 2.11.} A {\it braided Lie derivation} $\L{\l}$ with respect to element
$\l\in\N{k}$ will mean the braided derivation
$$\L{\l} =[\de ,\imath_{\l}]_{\hat\s}.$$
Note that $\L{\l}\in D_{k+1}(\M ),$ if $\l\in\N{k} .$

Below we collect together the main properties of braided Lie derivations.
\proclaim{Theorem} The braided Lie derivations are $\hat\s$--derivations of
the $\hat\s$--commutative algebra $\M$ of braided differential forms such that:
\roster
\item $[\de ,\L{\l}]_{\hat\s}=0,\quad\forall\l\in\N* ,$
\item $h(\L{\l} )=\L{h(\l )},\quad\forall h\in H,$
\item The braided commutator $[\L{\l_1},\L{\l_2}]_{\hat\s}$ is a braided Lie
derivation $\L{\{\l_1,\l_2\}}$ for some element $\{\l_1,\l_2\}$. This
element is called a braided differential Frolicher--Nijenhuis bracket.
\item Any $\hat\s$--derivation $f:\M\lra\M$ of the algebra braided differential
 forms $\M$ may be represented as follows:
$$f=\imath_{\l_1}+\L{\l_2},$$
for some uniquely determined elements $\l_1$ and $\l_2$.

Hence, there is a decomposition:
$$D_*(\M )={D_*}^{\text{alg}}(\M )\oplus {D_*}^{\text{Lie}}(\M ),$$
such that
$$\bigl[ {D_*}^{\text{Lie}}(\M ),{D_*}^{\text{Lie}} (\M )\bigr]_{\hat\s}
\subset {D_*}^{\text{Lie}}(\M ),$$
$$\bigl[ {D_*}^{\text{alg}}(\M ),{D_*}^{\text{alg}}(\M )\bigr]_{\hat\s}
\subset {D_*}^{\text{alg}}(\M ),$$
and
$$\bigl[\de ,{D_*}^{\text{alg}}(\M )\bigr]_{\hat\s}\subset {D_*}^{\text{Lie}}
(\M ),\qquad\bigl[\de ,{D_*}^{\text{Lie}}(\M )\bigr]_{\hat\s}=0.$$
\item The following braided analog of the infinitesimal Stokes theorem holds:
$$\left\{\aligned
&[\imath_{\l_1},\L{\l_2}]_{\hat\s}=\imath_{\{\l_1,\l_2\}} +(-1)^l
\L{\l_1\t\l_2},\\
&[\L{\l_2},\imath_{\l_1}]_{\hat\s}=(-1)^{l+1}\imath_{\{\l_2,\l_1\}} +
(-1)^{kl}\L{\sum \phi^{k+1}\s '(\l_1)\t\phi^{-l}\s ''(\l_2)}\\
\endaligned\right.$$
\endroster
where $\l_1\in\N{k} ,\l_2\in\N{l} .$
\endproclaim
\demo{Proof} Properties (1)--(3) of the braided Lie derivations are
consequences of properties 2.2. of braided derivations. We should remark only
that $\de$ is an $H$--invariant braided derivation.

To prove (4), we should note that $\l_2$ is determined by the restriction
$f\vert_A$. Then $f-\L{\l_2}$ is an algebraic braided derivation, and
therefore $f-\L{\l_1}=\imath_{\l_2}.$

Note also, that $[\de ,f]_{\hat\s}=\L{\l_2}$.

To prove (5), we decompose $[\imath_{\l_1},\L{\l_2}]_{\hat\s}$ as above:
$$[\imath_{\l_1},\L{\l_2}]_{\hat\s}=\imath_x+\L{y},$$
for some elements $x,y\in\N{*}.$
To determine $y$, we should look at the restriction $[\imath_{\l_1},\L{\l_2}]_
{\hat\s}$ on $A.$

One gets
$$[\imath_{\l_1}.\L{\l_2}]_{\hat\s}(a)=\imath_{\l_1}\circ\L{\l_2}(a)=
(-1)^{l+1}\imath_{\l_1}\circ\imath_{\l_2}(\de a)=(-1)^{l+1}
(\l_1\t\l_2)(\de a).$$
Therefore, $y=(-1)^{l+1}\l_1\t\l_2.$

Moreover, one has
$$\L{x} =[\de ,\imath_x]_{\hat\s}=[\de ,[\imath_{\l_1},\L{\l_2}]_{\hat\s}]_
{\hat\s}=[\L{\l_1},\L{\l_2}]_{\hat\s}=\L{\{\l_1,\l_2\}}.$$
In a similar one proves the second relation.
\qed\enddemo
{\bf 2.12.} In this secxtion we define a Lie structure on the modules of
braided
derivations. There are several definitions of braided Lie algebras (
[Gu],[Mj]).

All of them are based on the translation of the Jacobi identity into the
framework of braided categories.

Here we are suggesting to change our paradigm and to consider a braided Lie
coalgebra structure in modules of braided differential forms instead of Lie
algebra structure in modules of braided derivations. This makes it possible
to preserve some analogue of the skew symmetry property and write down Jacobi
identity as a braided version of the Master equation.

We should point out that our approach is based on the definition of Lie
coalgebra structures as  invariant braided derivations and we
therefore may exploit the theory of braided derivations developed in this
section.

Let $A$ be a $\s$--commutative algebra in the category $\C $ and let $M$ be
an $A-A$ bimodule. We fix some differential prolongation $\hat\s$ determined
by some central group--like element $\phi\in H.$

We will say that $M$ is a $\hat\s$--{\it symmetric bimodule} if $M_{(1)}$ is a
$\hat\s$--symmetric $\M$--bimodule.

In other words $M$ is a $\hat\s$--symmetric bimodule if
$$a\,m=\sum\s '(m)\cdot\phi^{-1}\s ''(a),$$
and
$$m\,a=\sum\phi\s '(a)\cdot\s ''(m),$$
for all $a\in A,\quad m\in M.$

Denote by $\ls{*} =\sum_{n\in\Bbb N}\ls{n}$ the $\hat\s$--commutative $\Bbb
N$--
graded algebra generated by $\ls{0} =A$ and $\ls{1} =M.$ Let $\w$ be a product
 in the algebra.

As above we denote by $D^{\text{alg}}_*(\ls{*})=\sum_{k\ge -1}D^{\text{alg}}_k
(\ls{*})$ the module of algebraic braided derivations of the algebra. Here we
denote by $D^{\text{alg}}_k(\ls{*})$ module of $\hat\s$--derivations
$f:\ls{*}\lra\ls{*}$ of degree $k\ge -1,$  such that $f\vert_A=0.$

Let $$\Ni{k} =Hom_{\hat\s}(M,\ls{k+1})$$ be a {\it Nijenhuis module} and
$$\Ni{*} =\sum_{k\ge -1}\Ni{k} $$ be a {\it Nijenhuis algebra} of the
bimodule $M.$

As we saw above, there is an isomorphism between $\Ni{k} $ and $D^{\text{alg}}
_k(\ls{*})$, given by inner braided derivations: $\a\in\Ni{k} \mapsto\imath
_\a\in D^{\text{alg}}_k(\ls{*}).$

We have two algebraic structures in $\Ni{*} :$
\roster
\item an associative graded algebra structure
$$\t :\Ni{k}\ot\Ni{l}\lra\Ni{k+l} ,$$
$$\a\ot\be\mapsto\a\t\be ,$$
where $$(\a\t\be )(m )=\imath_\a (\be (m )),$$
for all $m\in M,$ and
\item an algebraic Nijenhuis bracket
$$[.,.]:\Ni{k}\ot\Ni{l}\lra\Ni{k+l},$$
$$\a\ot\be\mapsto [\a ,\be]_{\hat\s},$$
where
$$\imath_{[\a ,\be ]_{\hat\s}}=[\imath_\a ,\imath_\be ]_{\hat\s},$$
or
$$[\a ,\be ]_{\hat\s}=(\a\t\be )-(-1)^{kl}\sum\phi^k\s '(\be )\t\phi^{-l}\s ''
(\a ).$$
\endroster
\example{Example}Let $k=-1.$ Then $\a\in\Ni{-1}=M^*$ determines the braided
inner derivation or a {\it braided annihilator operator} $\imath_\a :\ls{n}
\lra\ls{n-1}.$ Because $\Ni{-2}=0$ we get the following commutative relations:
$$[\imath_\a ,\imath_\be ]_{\hat\s}=0,$$
for all $\a ,\be\in\Ni{-1}.$
\endexample
\definition{2.13.Definition} We say that a $\hat\s$--derivation $\n$ of the
$\hat\s$-commutative algebra $\ls{*}$ of degree 1 is a {\it braided Lie
coalgebra structure} on the bimodule $M$ if
\roster
\item $\n$ is an $H$--invariant $\hat\s$--derivation,
\item the {\it Master equation }
$$[\n ,\n ]_{\hat\s}=0,\tag1$$
\endroster
holds.
\enddefinition
Therefore a Lie coalgebra structure determines two morphisms
\roster
\item a twisted derivation $\n :A\lra M,$ and
\item a braided symmetric co-product $\n :M\lra\ls{2},$
\endroster
such that the Master equation holds.

Our main example of a Lie coalgebra structure is the following
\proclaim{Theorem} The algebra braided differential forms $\M$ over $\s$--
commutative algebra $A$ is a braided Lie coalgebra with respect to the braided
differential $\de .$
\endproclaim
\demo{Proof}$[\de ,\de]=2\de^2=0.$
\qed\enddemo
{\bf 2.14.} Let $\n$ be a braided Lie coalgebra structure on a bimodule $M.$
Then we can define a $\n$--{\it Lie derivation} $\L{\a}^\n :\ls{*}\lra\ls{*}$
for any $\a\in\Ni{k},$ as above: $\L{\a}^\n =[\n ,\imath_\a ]_{\hat\s}\in D_
{k+1}(\ls{*}).$
\proclaim{Theorem} The braided $\n$-Lie derivations are $\hat\s$--derivations
of the $\hat\s$--commutative algebra $\ls{*}$ such that:
\roster
\item $[\n ,\L{\a}^\n ]_{\hat\s}=0,\quad\forall\a\in\Ni{*}.$
\item $h(\L{\a}^\n )=\L{h(\a )^\n},\quad\forall h\in H.$
\item The braided commutator $[\L{\a}^\n ,\L{\be}^\n ]_{\hat\s}$ is a braided
$\n$--Lie derivation $\L{\{\a ,\be\}^\n}$ for some element $\{\a ,\be\}^\n .$
The element is called a braided $\n$--differential Frolicher--Nijenhuis
bracket.
\endroster
\endproclaim
\demo{Proof} See the proof of Theorem 2.10.
\qed\enddemo
\head {\bf 3.Quantizations}\endhead
In this chapter we define quantizations of monoidal categories and functors
between them. We suggest two descriptions of quantizations: one in terms of
{\it quantizers} (and nonlinear cohomologies linked with the description),
and the other in terms of {\it Hochschild cohomologies of Grothendieck ring}
of the given monoidal category.

We show that quantizations "deform" all algebraic and differential objects
in the category and produce in this way quantizations of the braided
differential
operators.

{\bf 3.1.} Let $\C$ be a monoidal category equipped with
\roster
\item a bifunctor of tensor product
$$\ot :\C\times\C\lra\C ,$$
\item an associativity constraint
$$\a =\a_{X,Y,Z}:X\ot (Y\ot Z)\lra (X\ot Y)\ot Z ,$$
\item a unit object $k=k_\C$ with natural isomorphisms
$$\eta^l:k\ot X\lra X,\qquad\eta^r:X\ot k\lra X,$$
\endroster
such that the following MacLane coherence conditions hold [McL]:

(i) the pentagon axiom:
$$\CD
X\ot (Y\ot (Z\ot T))@>\a>>(X\ot Y)\ot (Z\ot T)@>\a>>((X\ot Y)\ot Z)\ot T\\
@A{id\ot\a}AA                @.                       @AA{\a\ot id}A\\
X\ot ((Y\ot Z)\ot T)@= X\ot ((Y\ot Z)\ot T) @>{\a}>>(X\ot (Y\ot Z))\ot T
\endCD$$
\comment
$$\diagram
X\ot (Y\ot (Z\ot T))&\rTo^\a&(X\ot Y)\ot (Z\ot T)&\rTo^\a&((X\ot Y)\ot Z)\ot
T\\
\dTo^{id\ot\a}& & & &\uTo_{\a\ot id}\\
X\ot ((Y\ot Z)\ot T)&&\rTo^\a&&(X\ot (Y\ot Z))\ot T\\
\enddiagram$$
\endcomment

(ii) the unity axiom:
\comment
$$\diagram
(X\ot k)\ot Y&\rTo\a&X\ot (k\ot Y)\\
\rdTo{\eta^r\ot id}&   &\ldTo{id\ot\eta^l}\\
&X\ot Y&\\
\enddiagram$$
\endcomment
$$(id\ot\eta^l)\circ\a_{X,k,Y}=\eta^r\ot id.$$
{\bf 3.2.} Let $\A$ and $\B$ be monoidal categories. We will denote the tensor
product in $\A$ by $\ot$ and that in $\B$ by $\hat\ot$ and associativity and
units constraints by $\a_\A ,\eta_\A ,$ and by $\a_\B ,\eta_\B$ respectively.

Let $\F :\A\lra\B$ be a unit preserving functor.
\definition{Definition}
\roster
\item We say that a natural isomorphism $\Q$
$$\Q_{X,Y}:\F (X)\hat\ot\F (Y)\lra\F (X\ot Y),\qquad X,Y\in\Cal Ob(\A ),$$
is a {\it quantization} of the functor $\F$ if $\Q$ preserves a unit object:
\comment
$$\diagram
\F (X)\hat\ot k_\B&\rTo{\Q_{X,k}}&\F (X\ot k_\A)\\
\rdTo{{\eta^r}_\B}&     &\ldTo{\F ({\eta^r}_\A )}\\
&\F (X),&\\
\enddiagram
\qquad\diagram
k_\B\hat\ot\F (X)&\rTo{\Q_{k,X}}&\F (k_{\A}\ot X)\\
\rdTo{{\eta^l}_\B}&    &\ldTo{\F {(\eta^l}_\A )}\\
&\F (X),&\\
\enddiagram
$$
\endcomment
$$\F (\eta^r_\A )\circ\Q_{X,k}=\eta^r_\B,\qquad\F (\eta^l_\A )\circ\Q_{k,X}
=\eta^l_\B ,$$
and the following diagram
$$\CD
\F (X)\hat\ot (\F (Y)\hat\ot\F (Z))@>{id\hat\ot\Q}>>\F (X)\hat\ot\F (Y\ot
Z)@>\Q>>\F (X\ot (Y\ot Z))\\
@V{\a_\B}VV                                                   @.
     @VV{\F (\a_\A )}V\\
(\F (X)\hat\ot\F (Y))\hat\ot\F (Z)@>{\Q\hat\ot id}>>\F (X\ot Y)\hat\ot\F
(Z)@>\Q>>\F ((X\ot Y)\ot Z)
\endCD$$
commutes.
\item We say that $\Q$ is a {\it quantization of monoidal category} $\A$ if
$\Q$ is a quantization of the identity functor $\F =id:\A\lra\A .$
\endroster
\enddefinition
 We show that the above commutative diagram is a realization of some type of
MacLane coherence conditions.

To do this, we introduce two monoidal categories. The first one is the
category [see McL] of binary words. The words are generated by two symbols:
$e_0$--empty word and (-)--placeholder. We convert the set of binary words
into a category $\Cal W$ by introducing one arrow between any binary words of
 the same length. It is a monoidal category under multiplication of
binary words with unit object $e_0.$

The second category is the monoidal category $\Cal It(\A ,\B )$ of iterates.
Objects of the category are pairs $(n,T),$ where $T:\A^n\lra\B$ is a functor.
Arrows in the category $f:(n.T)\lra (n,T')$ are natural transformations
$f:T\overset\t\to\lra T',$ [cf.McL]. We convert $\Cal It(\A ,\B )$ into a
monoidal category by introducing  multiplication $(m,S)\hat\ot (n,T)
=(n+m,S\hat\ot T),$ where $S\hat\ot T$ is the composite
$$S\hat\ot T:\A^{n+m}=\A^n\times\A^m\overset{S\times T}\to\lra\B\times\B
\overset{\hat\ot}\to\lra\B .$$
Any binary word $w$ of length $n$ determines a functor $\F_*(w):\A^n\lra\B$
obtained by replacing each placeholder (-) by the functor $\F .$

More precisely, the definition is given by recursion: if we determined
functors $\F_*(w_1)$ and $\F_*(w_2)$ for binary words $w_1,w_2$ of
respective lengths $n$ and $m,$ then $\F_*(w_1\cdot w_2)$ is determined by
the following diagram
$$\CD
\A^{n+m}=\A^n\times\A^m@>{\F_*(w_1\cdot w_2)}>>\B\\
@V{\F (w_1)\times\F (w_2)}VV    @AA{\hat\ot}A\\
\B\times\B@>\Q>>\B\times\B .
\endCD$$
\definition{Definition}[cf.Ep] We say that monoidal categories $\A$ and $\B$
are $\F$--{\it coherent} if $\F_*:\Cal W\lra\Cal It(\A ,\B )$ is a tensor
product preserving functor.
\enddefinition
The proof of the following theorem is analogous to the proof of
theorem 1.6.[Ep].
\proclaim{Theorem} Monoidal categories $\A$ and $\B$ are $\F$--coherent if
and only if $\Q$ is a quantization of the functor $\F .$
\endproclaim
There is an action of natural isomorphisms of the category $\A$ on the set of
all quantizations.

Let $\l :\A\overset\t\to\lra\A$ be a unit preserving natural isomorphism,
$\l_X:X\lra X,$ and let $\Q$ be a quantization of the functor $\F :\A\lra\B.$
Then we can build up a new isomorphism $\Q_\l :\F (X)\hat\ot\F (Y)\lra\F
(X\ot Y)$ by using the following commutative diagram
$$\CD
\F (X)\hat\ot\F (Y)@>{\Q_{X,Y}}>>\F (X\ot Y)\\
@V{\F (\l_X)\hat\ot\F (\l_Y)}VV  @V{\F (\l_{X\ot Y})}VV\\
\F (X)\hat\ot\F (Y)@>{(\Q_\l )_{X,Y}}>>\F (X\ot Y).
\endCD$$
Denote by $\Q (\F )$ the set of all the quantizations of the functor $\F$ and
by
$\Q H^2(\F )$ the set of equivalence classes with respect to the following
equivalence relation:
$$\Q\sim\Q '\Leftrightarrow\exists\text{\,a unit preserving transformation}
\,\l :\A\overset\t\to\lra\A ,\text{such that}\Q '=\Q_\l .$$
The set $\Q H^2(\F )$ is an analogue of the non--linear second cohomology
group associated to the functor $\F$ (see 3.6. below).

{\bf 3.3.} Quantizations transport all natural algebraic structures related
to tensor product.

Below we list the transport of main structures.

 {\it 1. Braidings and symmetries.}

Let $\F :\A\lra\B$ be a faithful functor and let $\s$ be a braiding or a
symmetry in a monoidal category $\A .$

Then the natural isomorphism $\s_\Q$, given by the commutative
diagram
$$\CD
\F (X)\hat\ot\F (Y)@>{\Q_{X,Y}}>>\F (X\ot Y)\\
@V{\s_\Q}VV                        @V{\F (\s )}VV\\
\F (Y)\hat\ot\F (X)@>{\Q_{Y,X}}>>\F (Y\ot X)
\endCD$$
is a braiding or, respectively, symmetry on the image of the functor $\F$
(see, for example, [L3]).

Therefore, if $\F$ is an isomorphism of categories then $\s_\Q$ is a braiding
 or a symmetry in the category $\B$.

{\it 2. Algebraic structures.}

Let $A$ be an algebra in a monoidal category $\A$ with multiplication law:
$\mu :A\ot A\lra A.$

Define a {\it quantization} $A_\Q$ of the algebra structure as the object
$A_\Q =\F (A)$ of the category $\B$ with new product
$$\mu_\Q =\F (\mu )\circ\Q_{A,A}:\F (A)\hat\ot\F (A)\lra\F (A).$$
It follows from the coherence conditions that the pair $(\F (A),\mu_\Q )$
determines an algebra structure in the category $\B .$

Indeed, the naturality of $\Q$ gives us the following commutative diagram
$$\CD
\F (A)\hat\ot\F (A\ot A)@>{id\hat\ot\F (\mu )}>>\F (A)\hat\ot\F (A)\\
@V{\Q_{A,A\ot A}}VV                              @V{\Q_{A,A}}VV\\
\F (A\ot (A\ot A))@>{\F (id\ot\mu )}>>\F (A\ot A).
\endCD$$
Therefore, for the proof of the associativity law we have
$$\align
\mu_\Q\circ (id\hat\ot\mu_\Q )&=\F (\mu )\circ\Q_{A,A}\circ (id\hat\ot\F (\mu
))
\circ\Q_{A,A}\\
&=\F (\mu )\circ\F (id\ot\mu )\circ\Q_{A,A\ot A}(id\hat\ot\Q_{A,A}).\\
\endalign$$
In the same way we get
$$\mu_\Q\circ (\mu_\Q\hat\ot id)=\F (\mu )\circ\F (id\ot\mu )\circ\Q_{A\ot A,A}
\circ (\Q_{A,A}\hat\ot id).$$
Analogously, if $A$ is a coalgebra in the category $\A$ with diagonal
$\D :A\lra A\ot A$ then $A_\Q =\F (A)$ is a coalgebra in the category $\B$
 with the new diagonal $\D_\Q =\Q_{A,A}^{-1}\circ\F (\D ).$

Let $\s$ be a braiding in the category $\A$ and let $(A,\mu )$ be an algebra
in $\A$. We define an algebra structure in $A^{\ot 2},$
$\mu^2_\s :A^{\ot 2}\ot A^{\ot 2}\lra A^{\ot 2}$ from the following
commutative diagram:
$$\CD
(A\ot A)\ot (A\ot A)@>{\mu^2_\s}>>A\ot A\\
@V{id\ot\s_{A,A}\ot id}VV          @\vert\\
(A\ot A)\ot (A\ot A)@>{\mu\ot\mu}>>A\ot A.
\endCD$$
\proclaim{Lemma} The morphism $\mu^2_\s$ determines an associative algebra
structure in  $\A .$
\endproclaim
\definition{Definition} [Mj]. A {\it braided} or $\s$--{\it bialgebra} in a
monoidal category $\A$ is an algebra $(A,\mu )$ and coalgebra $(A,\D )$
structures, such that $\D :A\lra A\ot A$ is an algebra homomorphism, where the
last algebra is considered with the structure $\mu^2_\s .$
\enddefinition
\proclaim{Theorem} Let $(A,\mu ,\D )$ be a $\s$--bialgebra. Then $(A_\Q ,\mu_
\Q ,\D_\Q )$ is a $\s_\Q$--bialgebra for any quantization $\Q .$
\endproclaim
 {\it 3. Module structures.}

Let $X$ be a left $A$--module in category $\A$ with multiplication $\mu^l:
A\ot X\lra X.$ {\it A quantization} of the module is the object $X_\Q =
\F (X)$ with the new product
$$\mu^l_\Q =\F (\mu )\circ\Q_{A,X}:\F (A)\ot\F (X)\lra\F (X).$$
One can show as above that $\mu_\Q$ determines a left $A_\Q$--module structure
 in the category $\B .$

In the obvious way we can apply the same procedure to right and bi--modules
structures.


{\bf 3.4.} Let $B$ be a bialgebra over $k .$  Quantizations of the monoidal
category $B-mod$ are determined by invertible elements $q\in B\ot B,$ such
that the following conditions
$$\align
(\D\ot id)(q)\cdot (q\ot 1)&=(id\ot\D )(q)\cdot (1\ot q),\tag1\\
(\e\ot id)(q)&=(id\ot\e )(q)=1,\tag2\\
q\cdot\D (b)&=\D (b)\cdot q,\forall b\in B,\tag3\\
\endalign$$
hold [L4].

Moreover, the action $\Q_{X,Y}:X\ot Y\lra X\ot Y$ is given by the formula
$$\Q_{X,Y}(x\ot y)=q\cdot (x\ot y).\tag4$$
 Analogously one can describe quantizations of the forgetful functor
$\F :B-mod\lra k-mod.$ Any such quantization is given by an element
$q\in B\ot B,$ but with conditions (1) and (2) only.
\definition{Definition} An element $q\in B\ot B$ is called a {\it quantizer}
of the bialgebra $B$ if conditions (1) and (2) hold.
\enddefinition
\proclaim{Theorem} Any braiding element $\s\in B\ot B$ of the bialgebra $B$ is
a
quantizer of the bialgebra. Moreover, $q=\s$ determines the quantization of
the category $B-mod.$
\endproclaim
\demo{Proof} Using conditions 1.3.(1) and (2), we get
$$\align
&(\D\ot id)(\s )\cdot (\s\ot 1)=\s_{23}\s_{13}\s_{12},\\
&(id\ot\D )(\s )\cdot (1\ot\s )=\s_{12}\s_{13}\s_{23},
\endalign$$
and therefore equation 3.4.(1) is a consequence of the Yang--Baxter equation.
\qed\enddemo

{\bf 3.5.} In this section we write down quantizations of the main structures
(see 3.3. above) in terms of quantizers.

{\it 1. Braidings.}
\proclaim{Theorem} Let $\s\in B\ot B$ be a braiding element and let
$q\in B\ot B$ be a quantizer of the category $B-mod$ (condition 3.4.(3))
holds).
Then
$$\s_q=q^{-1}\cdot\s\cdot\tau (q)$$
is a braiding element, too.
\endproclaim
\proclaim{Corollary} If $\s_1$ and $\s_2$ are braiding elements, then
$\s_1^{-1}\cdot\s_2\cdot\tau (\s_1)$ is a braiding element, too.
\endproclaim
{\it 2. Algebraic structures.}

Let $A$ be an algebra in the category $B-mod$ with multiplication $\mu :a_1\ot
a_2\mapsto a_1\cdot a_2.$

The quantization of $A$ is an algebra $A_q=A$ with new multiplication
$$a_1\underset{q}\to\cdot a_2=\sum q'(a_1)\cdot q''(a_2),$$
where $q=\sum q'\ot q''.$

We should remark that $A_q$ is a $B$--algebra if condition 3.4.(3) holds.

{\it 3. Module structures.}

Let $X$ be a left $A$--module in the category $B-mod$ with multiplication
$\mu^l:a\ot x\mapsto a\cdot x.$
The quantization is a left $A_q$--module $X_q=X$ with new multiplication
$$a\underset{q}\to\cdot x=\sum q'(a)\cdot q''(x).$$
In the same way we quantize right $A$--modules:
$$x\underset{q}\to\cdot a=\sum q'(x)\cdot q''(a).$$

{\bf 3.6.} Denote by $\Q (B)$ the set of all quantizers in bialgebra $B.$
\proclaim{Lemma} Let $U(B)\subset B$ be the group of units of the algebra
$B,$ such that

$\e (b)=1,\,\forall b\in U(B).$ Then the formula
$$b\in U(B),q\in\Q (B)\mapsto b(q)=\D (b)\cdot q\cdot b^{-\ot 2},$$
where $b^{-\ot 2}=b^{-1}\ot b^{-1},$ determines $U(B)$--action on $\Q (B).$
\endproclaim
\demo{Proof} We have
$$\align
&(\D\ot id)(b(q))\cdot (b(q)\ot 1)=\\
&(\D\ot id)\D (b)(\D\ot id)(q)(\D b^{-1}\ot b^{-1})(\D (b)\ot 1)(q\ot
1)(b^{-\ot 2}\ot 1)\\
&=(\D\ot id)\D (b)\,(\D\ot id)(q)\cdot (q\ot 1)\,b^{-\ot 3},\\
\endalign$$
and
$$\align
&(id\ot\D )(b(q))\cdot (1\ot b(q))=\\
&(id\ot\D )\D (b)(id\ot\D)(q)(b^{-1}\ot\D b^{-1})(1\ot\D (b))(1\ot q)(1\ot
b^{-\ot 2})\\
&=(id\ot\D )\D (b)(id\ot\D )(q)\,(1\ot q)\,b^{-\ot 3}.\\
\endalign$$
For conditions (2) we have
$$(\e\ot id)b(q)=b\,(\e\ot id)(q)\,\e (b)\,b^{-1}=\e (b)=1.$$
\qed\enddemo
Remark that in the case of forgetful functor $\F$ any unit preserving natural
transformation $\l$ from 3.2. has the form: $\l_X(x)=b\cdot x,\forall x\in X,$
and for some invertible element $b\in B,$ such that $\e (b)=1.$

 Denote by
$$\Q H^2(B)=\Q (B)/U(B)$$
the space of $U(B)$--orbits.

This space will be called the {\it non--linear cohomology} of the bialgebra
$B.$

To explain this definition let us consider a "linearization" of $\Q H^2(B).$

To do this, we fix a quantizer $q$ and describe the "tangent plane" $T_q\Q (B)$
at the point.

Let
$$q(t)=q+\sum_{i\ge 1}\a_i\,t^i$$
be a formal curve on $\Q (B),\quad\a_i\in B\ot B,\quad(\e\ot id)(\a )=(id\ot\e
)
(\a )=0.$

Substitute this expression in 3.4.(1) and look at coefficients of $t^k,\,k>0.$

We get the equations on $\a_i:$
$$\pa_q (\a_k)=\sum_{i+j=k,i\ge j,j>0}[\a_i,\a_j],\tag1$$
where
$$\align
[\a_i,\a_j]&=(\D\ot id)(\a_i)(\a_j\ot 1)-(id\ot\D )(\a_i)(\a_j\ot 1)\\
&+(\D\ot id)(\a_j)(\a_i\ot 1)-(id\ot\D )(\a_j)(1\ot\a_i),\tag2
\endalign$$
and the operator $\pa_q:B^{\ot 2}\lra B^{\ot 3}$ acts as follows
$$\align
\pa_q(\a )&= (\D\ot id)(q)(\a\ot 1)+(\D\ot id)(\a )(q\ot 1)\\
&-(id\ot\D )(q)(1\ot\a )-(id\ot\D )(\a )(1\ot q).\tag3
\endalign$$
Therefore, we have
$$T_q\Q (B)=ker\,\pa_q\bigcap ker (id\ot\e )\bigcap ker (\e\ot id ).$$
Any curve $b(t)=1+b\,t+....$ on the group $U(B)$ determines a curve
$$q(t)=\D (b(t))\,q\,b(t)^{-\ot 2}=q+t(\D (b)q-q(b\ot 1+1\ot b))+...$$
on the space $\Q (B).$

Note that $\e (b)=0.$

Therefore, tangent vectors to the orbit $U(B)q$ are elements of $Im\,\pa_q
\bigcap ker\,\e ,$ where $\pa_q:B\lra B^{\ot 2}$ is the following operator
$$\pa_q(b)=\D (b)q-q(b\ot 1+1\ot b).$$
It follows from the above constructions that the sequence
$$B\overset{\pa_q}\to\lra B^{\ot 2}\overset{\pa_q}\to\lra B^{\ot 3}$$
is a complex, and linearization of non--linear cohomology space may be
identified
with the second cohomology group $H^2(B,q)$ of the normalized complex.

We should note also that the complex above coincides with the standard complex
for Hochschild cohomology of the coalgebra at the point $q=1.$

 Let $\a\in B\ot B$ be a tangent vector to $\Q (B),$ i.e.
$$\pa_q(\a )=0,\qquad\text{and}\qquad (\e\ot id)(\a )=(id\ot\e )(\a )=0.$$
To build up a curve 3.5. $q(t)$, such that $\a_1=\a$ we need some additional
conditions on $\a .$

Namely, take $k=2$ in formula 3.5.(1).

We get
$$\pa_q(\a_2)=[\a ,\a].\tag1$$
It is easy to check that the condition
$$[\a ,\a]=0\,mod\,Im\,\pa_q$$
depends on the cohomology class of $\a$ only, and $[\a ,\a]\in Ker\,\pa_q,$
for all $\a\in ker\,\pa_q.$
Therefore, we obtain a bracket
$$[{},{}]_q:H^2(B,q)\ot H^2(B,q)\lra B^{\ot 3}/Im\,\pa_q.$$
This bracket will be called a $q$--{\it bracket.}

Hence, a vector $\bar\a\in T_q\Q H^2(B)$ is a tangent vector  to
some curve on $\Q H^2(B)$ if $[\bar\a ,\bar\a]_q=0.$
\definition{Definition} A $q$--{\it Poisson structure} on the bialgebra $B$
is a second cohomology class $\bar\a\in H^2(B,q),$ such that
$$[\bar\a ,\bar\a]_q=0.$$
\enddefinition
\example{Examples}

(1) Let us consider the category of $G$--graded modules over an Abelian
finite group $G$. Then conditions 3.4. mean that any quantizer $q$ is a
2--cocycle on the group with values in the unit group
$U(k):\quad q\in Z^2(G;U(k)).$

Therefore, in the given case : $\Q H^2(k(G))=H^2(G,U(k)).$

(2) Let $B$ be a commutative and co-commutative bialgebra over $k=\Bbb C.$
We consider quantizers of the {\it Moyal type:}
$$q=exp(\a ),$$
for some element $\a\in B\ot B.$

For this case the main equation 3.4.(1) takes the form:
$$\a\ot 1-(id\ot\D )(\a )+(\D\ot id)(\a )-1\ot\a=0.$$
The last equation means that $\a$ isa 2-cocycle on the {\it coalgebra} $B$
with values in $\Bbb C.$

For instance, if $B=\Cal D(\Bbb T^n)$ is the Hopf algebra of distributions
on $n$--dimentional torus $\Bbb T^n,$ and $\a$ is an invariant Poisson
structure on $\Bbb T^n,$ then the above formula gives the Moyal
quantization [BFFLS,V].

(3) Let $\imath :G_1\lra G$ be a subgroup of group $G.$ Then $\imath$
determines
a Hopf algebra homomorphism $\imath_*:k[G_1]\lra k[G]$ (or $\imath_*:\Cal
D(G_1)
\lra\Cal D(G)$ for the case of compact Lie groups), and $\imath_*(q)$ is a
quantizer on $G$, if $q$ is a quantizer on $G_1.$

Applying the remark to a maximal torus of a Lie group $G$, we obtain a class
of Moyal type quantizations [cf.R].
\endexample

{\bf 3.7.} Let $G$ be a semi--simple Lie group, $k=\Bbb C,$ and $\C$  a
monoidal category of finite dimensional $G$--modules over $\Bbb C.$

Denote by $K(G)$ the representation algebra of finite dimensional $G$--modules
 over $\Bbb C.$

In this section we show that quantizations of $\C$ may be described in
terms of multiplicative 2--cohomologies of $K(G).$

To do this, we introduce a new algebra $E(\hat G)$ generated by all formal
finite sums $f=\sum_{\g\in\hat G}f(\g )\,\g$, where $\hat G$ is the set of
all the finite dimensional irreducible representations of $G$, and
$f(\g )\in Hom_G(nX_\g ,nX_\g )\simeq Mat_n(\Bbb C),$ for some natural
$n\in\Bbb N.$

Here we denote by $X_\g$ a representative module of $\g .$

We convert $E(\hat G)$ into an algebra over $K(G)$ by introducing the
following operations:
$$f+g\overset\text{def}\to=\sum_{\g\in\hat G}(f(\g )\oplus g(\g ))\,\g ,$$
and
$$f\cdot g\overset\text{def}\to=\sum_{\g\in\hat G}(f(\g )\ot g(\g ))\,\g ,$$
with $K(G)$--action:
$$\a\cdot f\overset\text{def}\to=\sum_{\g\in\hat G}f(\g )\,\a\g=\sum_
{\d ,\g\in\hat G}(id\ot f(\g ))\vert_{\d}\,\d ,$$
where $\a\g =\a\ot\g =\sum_{\d\in\hat G}m^{\d}_{\a\g}\d ,$ and $\vert_\d$ is
the restriction on the $\d$--component, $m^{\d}_{\a\g}\in\Bbb N$ are the
multiplicities of $\d$ in the tensor product $\a\ot\g .$

Any quantization $\Q$ of the monoidal category determines morphisms
$$\om_{\a ,\be}=\Q_{X_\a,X_\be}:X_\a\ot X_\be\lra X_\a\ot X_\be$$
for all $\a ,\be\in K(G).$

We write down the morphisms in the form
$$\om_{\a ,\be}=\sum_{\g\in\hat G}\om_{\a ,\be}(\g )\,\g$$
where $\om_{\a ,\be}(\g )$ is the restriction of $\om_{\a ,\be}$ on the $\g$--
component.

Therefore, any quantization $\Q$ determines 2--cochain
$$\om :K(G)\times K(G)\lra E(\hat G),$$
with the additional condition
$$\om_{\a ,\be}(\g )\in Hom_G(n_{\a ,\be}^\g\,X_\g,
n_{\a ,\be}^\g\,X_\g)\simeq Mat_{n_{\a ,\be}^\g}(\Bbb C),\tag1$$
if $\a\be =\sum_{\g\in\hat G}n_{\a ,\be}^\g\,\g .$

Moreover, $\om$ is a normalized 2-cochain : $\om_{\a ,1}=\om_{1,\a}=1,$ and
commutativity of diagram 3.2. yields the following multiplicative 2-cocycle
property
$$\om_{\a ,\be\g}\cdot\a (\om_{\be ,\g})=\om_{\a\be ,\g}\cdot\g (\om_{\a,\be}).
\tag2$$
We say that $\om$ is a {\it multiplicative 2-cocycle} if (2) holds for all
$\a ,\be ,\g\in K(G).$

We say that $\om$ is a {\it restricted 2-cocycle} if $\om$ is a multiplicative
2-cocycle satisfying the additional condition (1).

Summarizing, we obtain the following
\proclaim{Theorem} Any quantization of the monoidal category of finite
dimensional $G$--modules is given by a multiplicative normalized and restricted
2-cocycle on the representation algebra with values in $E(\hat G).$
\endproclaim
\remark{Remark}Let $G$ be a compact Lie group. Then  in the same way one can
describe quantizations of the monoidal category of unitary modules.
\endremark

{\bf 3.8.} In this section we consider quantizations of the category such that
$\om_{\a ,\be}(\g )\in\Bbb C^*,$ for all $\a ,\be ,\g\in\hat G,$ and
 $$\om (\a ,\be )=exp (2\pi\imath\,\th (\a ,\be )),\tag1$$
for some 2-cochain $\th :K(G)\times K(G)\lra E(\hat G),$ where
$\th_{\a ,\be}(\g)\in\Bbb C$.

Let
$$d:C^k(K(G),E(\hat G))\lra C^{k+1}(K(G),E(\hat G))$$
be the Hochschild differential .

Then $\om_{\a ,\be}$ commute and therefore conditions 3.7.(2) take the form
$$d(\th )(\a ,\be ,\g )=\a (\th (\be ,\g ))-\th (\a\be ,\g )+\th (\a ,\be\g )-
\g (\th (\a ,\be ))\in\Bbb Z.\tag2$$
 Denote by $C^k_{\Bbb Z}(K(G),E(\hat G))\subset C^k(K(G),E(\hat G))$ the
integer subcomplex
and let $C^k_{\Bbb C/\Bbb Z}(K(G),E(\hat G))$ be the quotient complex.

Then condition (2) means that $\th$ is a 2-cocycle in the quotient complex.

Therefore, we may reformulate theorem 3.7. in the following way:
\proclaim{Theorem} Quantizations of type (1) are determined by Hochschild
2-cocycles of the quotient complex $C^{\t}_{\Bbb C/\Bbb Z}(K(G),E(\hat G)).$
\endproclaim
\example{Examples}

(1) Let us consider the category of $G$-modules over an Abelian finite group
$G$, $k=\Bbb C.$

Let $\hat G$ be the dual group. Then we have $X_\a\ot X_\be =X_{\a\be},$
fot all $\a ,\be\in\hat G.$

Therefore, $\om_{\a ,\be}=\tilde\om_{\a ,\be}\cdot (\a\be ),$ where
$$\tilde\om :\hat G\times\hat G\lra\Bbb C^*$$
is a 2-cocycle.

(2) Applying the same construction to the category of finite dimentional
$\Bbb T^n$--modules, we get the Hochschild 2-cocycle:
$$\tilde\om :\Bbb Z^n\times\Bbb Z^n\lra\Bbb C^*.$$

(3) The same construction may be applyed for the case of arbitrary compact
Lie group $G$, if we consider the "1-particle" interaction:
$\om_{\a ,\be}=\tilde\om_{\a\t\be}(\a\t\be ),$
where $\a\t\be$ is the highest part of the representation $\a\be .$

(4) Let $k=\Bbb C$ and $G=S_3$ be the permutation group on three elements.
The representation algebra $k(S_3)$ is generated by $x_1$--non-trivial
1-dimentional irreducible  represantation, and $x_2$--2-dimentional
irreducible representation with the following relations:
$$x_1^2=1,\quad x_1x_2=x_2,\quad x_2^2=x_2+x_1+1.$$
Set $\om_{ij}=\om_{x_ix_j}.$

It easy to find that
$$\om_{11}=a^2\cdot 1,\quad\om_{12}=\om_{21}=a\cdot x_2,\quad\om_{22}=ab\cdot
 1+b\cdot x_1+c\cdot x_2,$$
for some $a,b,c\in\Bbb C^*.$

We have the following action of element $h=h_1\cdot x_1+h_2\cdot x_2$ on $\om$:
$$h(\om_{11})=h_1^{-2}\om_{11},\quad h(\om_{12})=h_1^{-1}\om_{12},$$
and
$$h(\om_{22})=h_2^{-1}c\cdot x_2+h_1h_2^{-2}b\cdot x_1+h_2^{-2}ab\cdot 1.$$
Therefore, we have 1-parameter family of quantizations, considered up to
trivial.
\endexample


{\bf 3.9.} Let $\C$ be a monoidal category $B-mod$ for some $k$--bialgebra
$B$ and let $\C_0$ be a monoidal category of $k-mod.$

In this section we describe quantizations of the category $\G$ of $\Bbb N$--
graded objects over $\C$ and quantizations of the forgetful functor
$\F :\G\lra\Cal Gr(\C_0).$

Let $\hat\Q$ be a quantization of the forgetful functor. The quantization is
given by the quantizer $\{q_{n,m}\} ,$ where
$$\hat\Q_{X,Y}(x_n\ot y_m)=q_{n,m}\cdot (x_n\ot y_m),\tag0$$
and $q_{n,m}\in B\ot B,\quad n,m\in\Bbb N.$

Suppose that $q_{0,0}=q$ is a quantizer of the bialgebra.

Now equations 3.4.(1) and 3.4.(2) take the form:
$$\align
(id\ot\D )(q_{n,m+k})\cdot (1\ot q_{m,k})&=(\D\ot id)(q_{n+m,k})\cdot
(q_{n,m}\ot 1),\tag1\\
(\e\ot id)(q_{0,n})&=(id\ot\e )(q_{n,0})=1.\tag2
\endalign$$
Apply the morphism $id\ot\e\ot id $ to both sides of (1).

We get
$$q_{n,m+k}\cdot (1\ot p_{m,k})=q_{n+m,k}\cdot (\bar p_{n,m}\ot 1),\tag3$$
where $p_{m,k}=(\e\ot id)q_{m,k},\quad\bar p_{m,n}=(id\ot\e )q_{n,m}.$

It follows from (2) that $p_{0,k}=1,\quad\bar p_{n,0}=1.$

Apply $\e\ot id$ and $id\ot\e$ to equation (3).

We get
$$\align
p_{n,m+k}\cdot p_{m,k}&=p_{m+n,k}\cdot\e (\bar p_{n,m}),\tag4\\
\bar p_{n,m+k}\cdot\e (p_{m,k})&=\bar p_{n+m,k}\cdot\bar p_{n,m}.\tag5
\endalign$$
Let
$$f_{n,m}=\e (p_{n,m})=\e (\bar p_{n,m}).$$
By applying counit $\e$ to equations (4) or (5), we get
$$f_{n,m+k}\cdot f_{m,k}=f_{m+n,k}\cdot f_{n,m}.$$
It follows that $f_{n,m}$ determines a 2-cocycle  on the group $\Bbb Z$ and
therefore
$$f_{n,m}=\frac{f(n)f(m)}{f(n+m)}$$
for some function $f:\Bbb Z\lra U(k).$

This means that up to equivalence we may suppose that $f_{n,m}\equiv 1.$

In the last case equations (4) and (5) take the form
$$\left\{\aligned
p_{n,m+k}\cdot p_{m,k}&=p_{n+m,k},\quad p_{0,k}=1,\\
\bar p_{n+m,k}\cdot\bar p_{n,m}&=\bar p_{n,m+k},\quad\bar p_{n,0}=1,
\endaligned\right.$$
It follows from the system that
$$p_{n,m}=g(n+m-1)g(m-1)^{-1},$$
for some function $g:\Bbb Z\lra A,$ and analogously
$$\bar h_{n,m}=h(n+m-1)h(n-1)^{-1},$$
for some function $h:\Bbb Z\lra A.$

Substitute this expressions in equation (3).

We get $h\equiv 1,\quad g\equiv 1$
and $q_{n.m}=Q(n+m-1),$ for some function $Q:\Bbb Z\lra A\ot A.$

Now it follows from equation (1) that $Q$ is a constant function.

Summarizing, we obtain the following
\proclaim{Theorem} Any quantization $\hat\Q$ of the forgetful functor
$\F :\Cal Gr(B-mod)\lra\Cal Gr(k-mod)$ is given by formula (0), where
$$q_{n,m}=\frac{f(n)f(m)}{f(n+m)}\cdot q$$
for some function $f:\Bbb Z\lra U(k)$ and quantizer $q.$

Any two  quantizations with given quantizer $q$ are equivalent.
\endproclaim
The proof of the theorem shows that the same result also holds for
quantizations of the category.

{\bf 3.10.} In this section we apply the above quantization procedure to the
modules of internal homomorphisms in the monoidal category $H-mod,$ where
$H$ is a Hopf algebra.

To do this, we remark that the composition $$f\in Hom(Y,Z),\,g\in Hom(X,Y)
\mapsto f\circ g\in Hom(X,Z)$$ defines an associative partially determined
product in the totality of  all internal homomorphisms.

Moreover, we shall identify elements of modules $x\in X$ with internal
homomorphisms $x:k\lra X,$ where $x:1\mapsto x.$

In terms of this identification, the evaluation map
$$x\in X,\,f\in Hom(X,Y)\mapsto y=f(x)\in Y$$
is the product $f\circ x.$

Let $\Q$ be a quantization of the category, given by the quantizer
$q\in H\ot H.$

 We define a new composition law $f\underset{q}\to\circ g$ as above
$$f\underset{q}\to\circ g=\sum q'(f)\circ q''(g).$$
For evaluation morphism we have
$$f_q(x):=f\underset{q}\to\circ x=\sum q'(f)(q''(x))=\sum q'_{(1)}
(f(Sq'_{(2)})q''(x)).$$
Note that $f\underset{q}\to\circ g=f\circ g$ if $f$ or $g$ is a morphism of
the category.

As above new composition defines an associative partially determined product
on the totality of all the internal homomorphisms.

{\bf 3.11.} Here we apply the above procedure of quantizations of internal
homomorphisms to modules of braided differential operators.

We start with the bimodule case. Let $\s$ be a braiding and $A$  a $\s$--
commutative algebra in the category. Let $X$ be an $A-A$  bimodule.

Let $\d^r_a$ and $\d^l_a$ be the $\d$--operations in the bimodule $X$, and
$$\align
&\d^l_{q,a}(x)=a\dq x-\sum\s'_q(x)\dq\s''_q(a),\\
&\d^r_{q,a}(x)=x\dq a-\sum\s'_q(a)\dq\s''_q(x),
\endalign$$
$\d$--operations in the $A_q-A_q$--bimodule $X_q.$
\proclaim{Lemma}One has
$$\d^l_{q,a}(x)=\sum\d^l_{q'(a)}(q''(x)),$$
and
$$\d^r_{q,a}(x)=\sum\d^r_{q''(a)}(q'(x)).$$
\endproclaim
\demo{Proof} We prove, for example, the first equality.
One has
$$\align
&a\dq x-\sum\s'_q(x)\dq\s''_q(a)=\\
&\sum q'(a)\cdot q''(x)-\sum q'\s'_q(x)\cdot q''\s''_q(a)=\\
&\sum q'(a)\cdot q''(x)-\sum\s' q''(x)\cdot\s'' q'(a)=\\
&\sum\d^l_{q'(a)}(q''(x)).
\endalign $$
\qed\enddemo
\proclaim{Corollary} There exist embeddings $X_\s\subset(X_q)_{\s_q}.$
\endproclaim
Applying the result to modules of braided differential operators, we obtain
the following
\proclaim{Theorem} For any braided $\s$--differential operator $\n\in Diff^\s_*
(X,Y)$ internal homomorphism $\n_q,$ where $\n_q(x)\overset\text{def}\to =\n
\cq x,$ is a braided $\s_q$--differential operator.

The correspondence $\n\mapsto\n_q$ determines morphisms
$$\hat q:Diff^\s_*(X,Y)\lra Diff^{\s_q}_*(X_q,Y_q)$$
of modules of braided differential operators.

The morphism preserves the composition $\n_1\circ\n_2\mapsto (\n_1)_q
\cq (\n_2)_q$ and the order of braided differential operators.
\endproclaim
\definition{Definition} The operator $\n_q\in Diff^{\s_q}_i(X_q,Y_q)$ will be
called a {\it quantization} of the operator $\n\in Diff^{\s}_i(X,Y).$
\enddefinition

{\bf 3.12.} Let $\n\in Der(A,X)$ be a $\s$--derivation of algebra $A$ with
values in a left $A$--module $X.$

Then $\n_q(1)=\n (1)=0,$ and therefore $\n_q$ is a derivation:
$\n_q\in Der(A_q,X_q).$

Applying the quantization to the representative objects, we obtain a morphism
of braided differential forms
$$\hat q:\O{1}{A}\lra\O{1}{A_q}.$$
Note that $H$--invariance of the differential $d$ gives us $d_q=d,$ and
therefore morphism $\hat q$ has the form:
$$\hat q:a\,db\in\O{1}{A}\mapsto a\dq db\in\O{1}{A_q}.$$
Moreover,we can define in the same way morphisms
$$\hat q_\phi:\O{1}{A,\phi}\lra\O{1}{A_q,\phi},$$
where
$$\hat q_\phi :a\,d_\phi b\mapsto a\dq d_\phi b,$$
for any differential prolongation of $\s$ is given by the element $\phi .$

In an obvious way the morphism may be extended to a homomorphism of $\hat\s$--
commutative algebra $\O{*}{A,\phi}$ into the $\hat\s_q$--commutative algebra
$\O{*}{A_q,\phi}.$
\proclaim{Theorem}A quantization generates morphism of braided differential
forms
$$\hat q_\phi :\O{*}{A,\phi}\lra\O{*}{A_q,\phi}$$
such that
\roster
\item $\hat q_\phi$ is a morphism of a $\hat\s$--commutative algebra into a
$\hat\s_q$--commutative algebra, considered with respect to
$\underset{q}\to\w$ multiplication,
\item $\hat q_\phi\circ d_\phi =d_\phi\circ\hat q_\phi .$
\endroster
\endproclaim
Denote by $\O{*}{A,\phi}_q=\sum_{i>0}\O{i}{A,\phi}_q$ the kernel of
$\hat q_\phi$. It is an ideal of $\hat\s$--commutative algebra closed with
respect to $d_\phi .$

 A {\it Quantum cohomology kernel} $H^i_q(A,\phi )$ of
the $\s$--commutative algebra $A$ and a quantization $q$ is the cohomology of
the complex
$$\O{1}{A,\phi}_q\overset{d_\phi}\to\lra\O{2}{A,\phi}_q\overset{d_\phi}\to
\lra\cdots\lra\O{i}{A,\phi}_q\overset{d_\phi}\to\lra\O{i+1}{A,\phi}_q\lra
\cdots .$$

\enddocument